\documentclass[journal,twoside,web]{ieeecolor}
\usepackage{generic}
\usepackage{cite}
\usepackage{amsmath,amssymb,amsfonts}
\usepackage{graphicx}
\usepackage[inkscapearea=page]{svg}
\usepackage{caption}
\usepackage{subcaption}
\usepackage{textcomp}
\usepackage{booktabs}
\usepackage{algorithm}
\usepackage{color}
 
\usepackage{hyperref}
\usepackage{balance}
\usepackage{pgfplots} 
\pgfplotsset{compat=1.11}

\usepackage{algpseudocode}
\def\BibTeX{{\rm B\kern-.05em{\sc i\kern-.025em b}\kern-.08em
    T\kern-.1667em\lower.7ex\hbox{E}\kern-.125emX}}
\begin{document}
 
\title{Advancing Stuttering Detection via Data Augmentation, Class-Balanced Loss and Multi-Contextual Deep Learning}

\author{Shakeel A.~Sheikh, Md Sahidullah, Fabrice Hirsch, Slim Ouni
 \thanks{Shakeel A.~ Sheikh, Md Sahidullah, and Slim Ouni are from Universit\'{e} de Lorraine, CNRS, Inria, LORIA, F-54000, Nancy, France (e-mail: {shakeel-ahmad.sheikh, md.sahidullah, slim.ouni}@loria.fr).}
\thanks{Fabrice Hirsch is from Université Paul-Valéry Montpellier, CNRS, Praxiling, Montpellier, France (e-mail: fabrice.hirsch@univ-montp3.fr).}
}

\maketitle

\begin{abstract}
 Stuttering is a neuro-developmental speech impairment characterized by uncontrolled utterances (interjections) and core behaviors (blocks, repetitions, and prolongations), and is caused by the failure of speech sensorimotors. Due to its complex nature, stuttering detection (SD) is a difficult task. If detected at an early stage, it could facilitate speech therapists to observe and rectify the speech patterns of persons who stutter (PWS). The stuttered speech of PWS is usually available in limited amounts and is highly imbalanced. To this end, we address the class imbalance problem in the SD domain via a multi-branching (MB) scheme and by weighting the contribution of classes in the overall loss function, resulting in a huge improvement in stuttering classes on the SEP-28k dataset over the baseline (\emph{StutterNet}). To tackle data scarcity, we investigate the effectiveness of data augmentation on top of a multi-branched training scheme. The augmented training outperforms the MB \emph{StutterNet} (clean) by a relative margin of 4.18\% in macro F1-score ($\mathcal{F}_1$). In addition, we propose a multi-contextual (MC) \emph{StutterNet}, which exploits different contexts of the stuttered speech, resulting in an overall improvement of 4.48\% in $\mathcal{F}_1$ over the single context based MB \emph{StutterNet}. Finally, we have shown that applying data augmentation in the cross-corpora scenario can improve the overall SD performance by a relative margin of 13.23\% in $\mathcal{F}_1$ over the clean training.
\end{abstract}


\section{Introduction}
\label{sec:introduction}
\vspace{-0.15cm}

\IEEEPARstart{S}{peech} impairments, often known as speech disorders, are difficulty in producing speech sounds. These speech difficulties usually take the form of dysarthria, cluttering (poorly intelligible speech), lisping, apraxia, and stuttering~\cite{guitar2013stuttering, duffy2013motor, ratner2018fluency, ward2008stuttering, kehoe2006speech}. Only a few percentage (5-10\%) of the world population can produce accurate speech units, the rest encounter some type of speech disorder in their life span~\cite{hurjui2016spatial}.
Of these speech impairments, stuttering is the most predominant one~\cite{guitar2013stuttering}. Stuttering\footnote{Stuttering is also called stammering. In this paper, we will use the terms disfluency, stuttering, and stammering interchangeably.}, is a neuro-developmental speech disorder, in which the flow of speech is disturbed by abnormally persistent and involuntarily speech sounds, which usually take the shape of \textit{core behaviors:} including blocks, prolongations and repetitions~\cite{guitar2013stuttering}. Stuttering is complex and the several factors that lead to it are delayed childhood development, stress, and speech motor abnormalities~\cite{guitar2013stuttering}. In~\cite{smith2017stuttering}, Smith and Weber put forward the multifactorial dynamic pathway theory, where they argued that stuttering occurs due to the failure of the nervous system. People with stuttering (PWS) exhibit impairment in sensorimotor processes which are responsible for the production of speech, and its direction is influenced by emotional and linguistic aspects. 

In conventional stuttering detection (SD) and therapy sessions, the speech therapists or speech-language pathologists manually analyze the PWS’ speech~\cite{noth2000automatic}. The speech therapists observe and monitor the speech patterns of PWS to rectify them~\cite{guitar2013stuttering}. This convention of SD is very laborious and time-consuming and is also inclined toward the idiosyncratic belief of speech therapists. In addition, the automatic speech recognition systems (ASR) are working fine for normal fluent speech, however, they are unsuccessful in recognizing the stuttered speech~\cite{mitra2021analysis}, which makes it impractical for PWS to easily access virtual assistants like Apple Siri, Alexa, etc. As a result, interactive automatic SD systems that provide an impartial objective, and consistent evaluation of stuttering speech are strongly encouraged. The SD can also be used to adapt and improve ASR virtual assistant tools for stuttered speech. Despite the fact \textcolor{black}{that} having numerous potential applications, very little research attention has been given to the domain of SD and measurement. The detection and identification of stuttering events can be quite a difficult and complex problem due to several variable factors including language, gender, age, accent, speech rate, etc.
\par 
\textcolor{black}{The main goal of this work is to build robust automatic stuttering detection systems capable of detecting multiple stuttering types based on speech modality, which later on can be deployed as a tool in real-world settings by providing a means to both PWS and speech therapists to keep track of the stutter speech. These systems can later on be further improved by providing a feedback mechanism to PWS and help them to rectify their stuttering. }
\par 
A significant amount of work is done in the detection of other speech disorders~\cite{disorder} like dysarthria~\cite{dysarthric} and Parkinsons~\cite{quan}, but stuttering has not been addressed widely even though it is the most common one. In this paper, we propose a deep learning framework for robust SD. The automatic detection of stuttering can help in the treatment of stuttering if detected at an early age~\cite{guitar2013stuttering}.
\par 
Most of the computer-based SD methods are based either on ASR systems~\cite{Alharbi2018, ALHARBI2020101052} or language models~\cite{Zayats+2016, chen}. These methods are two-stage approaches that first convert the acoustic speech signals into their corresponding spoken textual modality, and then detect stuttering by the application of language models. Even though this ASR-based two-stage approach for identifying stuttering has shown promising results, the dependence on the ASR unit makes it computationally costly and prone to error. Moreover, the adaption towards the ASR task results in the possible loss of stuttering relevant information such as prosodic and emotional content.
\par 
 In recent decades, the applications of deep learning have grown tremendously in speech recognition~\cite{speechrecog}, speaker recognition~\cite{latentspeech}, speech synthesis~\cite{synthesis}, emotion detection~\cite{emotion}, voice conversion~\cite{conversion},  voice disorder detection~\cite{disorder} including Parkinson's disease detection~\cite{parkinson} and dysarthric speech detection~\cite{dysarthric, inadysarthric}. Inspired by the human auditory temporal mechanism, \textcolor{black}{Kodrasi \emph{et al.}}~\cite{inadysarthric} recently proposed a convolutional neural network based dysarthric speech detection method by factoring the speech signal into two discriminative representations including temporal envelope (stress, voicing, and phonetic information) and fine structure (breathiness, pitch, and vowel quality) and reported the state-of-the-art results in \textcolor{black}{dysarthria} detection. However, the application of deep learning in SD is limited. \textcolor{black}{The acoustic properties of speech disfluencies are different for different disfluencies which can help to discriminate from fluent voice. Due to the presence of these acoustic cues in the stuttered embedded speech, deep learning models
can be used to exploit these acoustic cues in the detection and identification of stuttering events.} Most of the SD existing methods employ spectral features including \emph{mel-frequency cepstral coefficients} (MFCCs) and spectrograms or their variants that capture the stuttering-related information. 

The earlier studies in this domain applied shallow deep learning approaches in SD. In 1995, Howell \emph{et al}.~\cite{howell1995automatic} employed two fully connected artificial neural networks \textcolor{black}{for the identification of two types of disfluencies, namely, repetition and prolongation}. They extracted autocorrelation features, envelope parameters, and spectral information and used these features as an input to the artificial neural networks for SD. The network was trained on 12 speakers with 20 autocorrelation features and 19 vocoder coefficients. In 2009, Ravikumar et al~\cite{ravikumar2008automatic} proposed multi-layered perceptron for the repetition type of stuttering. The network was trained by using MFCC input features on 12 different disfluent speakers. In 2019, B. Villegas \emph{et al}.~\cite{villegas2019novel} trained multi-layer perceptron on 10-dimensional respiratory features for the block SD. They used a dataset of 68 Latin American Spanish speakers in their case study. In a recent study, Kourkounakis \emph{et al}.~\cite{kourkounakis2020detecting} proposed \textcolor{black}{residual network and bi-directional long short-term memory (ResNet+BiLSTM)} based binary deep learning classifiers for the detection of six different types of disfluencies including prolongation, word repetition, sound repetition, phrase repetition, and false starts. They used spectrograms as input features and reported promising results on a small subset (25 speakers) from the UCLASS dataset~\cite{kourkounakis2020detecting}. In another study Lee \emph{et al}.~\cite{sep28k} \textcolor{black}{curated} a large stuttering dataset (SEP-28k) and utilized the \textcolor{black}{convolutional long short-term memory (ConvLSTM)} model for the detection and identification of six types of stuttering, namely, blocks,  prolongations, sound repetitions, word repetitions, and interjections. The model takes 40-dimensional input MFCCs, eight-dimensional articulatory features, 41-dimensional phoneme feature vector, and three-dimensional pitch features. Sheikh \emph{et al.}~\cite{sheikh:hal-03227223} recently proposed a single multi-class time delay neural network (TDNN) based \emph{StutterNet} classifier which is capable of detecting core behaviors and fluent speech segments and gives promising detection performance on a larg subset of UCLASS (100+) speakers compared to the state-of-the-art classifiers. The model solely takes 20-dimensional MFCC features as an input. In another recent study M. Jouaiti \emph{et al.}~\cite{melanie} introduced phoneme-based \textcolor{black}{bi-drectional long short-term memory (BiLSTM) model} for SD by mixing the SEP-28k and UCLASS datasets. The model is trained on 20-dimensional MFCCs and 19-dimensional phoneme input features. The disfluencies considered in this work are prolongations, repetitions, and interjections. A detailed summary and comparison of various feature extraction methods and classifiers can be found in~\cite{sheikh2021machine}.
 \par
\textcolor{black}{This work provides a complete deep learning framework for robust SD following our preliminary initial investigations~\cite{sheikh:hal-03227223} in which \emph{StutterNet} yields state-of-the-art SD results. In this study, we identify the limitations of \emph{StutterNet} and proposed further advancements to address those drawbacks. Our main contributions are summarized below:
   \begin{itemize}
    \item \emph{Solution for class imbalance problem}: The standard stuttering datasets suffer from class imbalance problems. We introduced two strategies based on weighted loss and multi-branch architecture to tackle the class imbalance problem.
    \item \emph{Introducing data augmentation}: The stuttering datasets have a limited amount of training data and this makes it difficult to apply advanced deep learning models with a large number of parameters. To address this limitation, this work introduces audio data augmentation. To the best of our knowledge, this is the first work to apply audio data augmentation in the SD problem.
    \item \emph{Introducing multi-contextual architecture}: Stuttering detection is a special type of speech characterization problem where each class (i.e., stuttering types) has a varying duration. For example, block lasts for a shorter duration than prolongations and repetitions. Therefore, a fixed length of context in the basic \emph{StutterNet} framework might not be the optimized one for detecting all types of stuttering. We introduce a multi-contextual architecture that uses different context lengths in a parallel fashion.    
    \end{itemize}  
}

The remaining of the paper is organized as follows. Section~\ref{stutternetoverview} describes the \emph{StutterNet} and analyzes its deficiencies. Section~\ref{address} discusses the proposed methodology for addressing the deficiencies. Section~\ref{expts} details the experimental design, metrics used and datasets. Section~\ref{results} discusses the experimental results on class balanced, data augmentation, MC~\emph{StutterNet} and cross corpora scenario. Finally, in Section~\ref{conc}, we conclude with the possible future directions.

\vspace{-0.1cm}
\section{StutterNet: overview \& limitations}
\label{stutternetoverview}
\subsection{StutterNet}
\vspace{-0.1cm}
Most of the earlier work employed only a small set of disfluent speakers in their experimental studies, and has approached the SD problem as a binary classification problem: disfluent vs fluent identification or one vs other type~\cite{sheikh2021machine}. The \emph{StutterNet} we propose in our earlier work, is a time delay neural network based architecture that has been used to tackle the SD as a multi-class classification problem. The \emph{StutterNet} takes 20 MFCCs as input features with a frame length of 20~ms, mean-normalized with a hop length of 10~ms. Usually, the initial layers in a standard deep neural network learn wider contexts when processing a temporal input signal. However in \emph{StutterNet} network as shown in Table \ref{tab:snet}, the initial layers learn and capture only smaller contexts and the deeper ones compute the activations from a wider context, thus the deeper/higher layers are able to capture and learn the longer temporal contexts. The network consists of five time delay layers with the first three focusing on $[t-2,~t+2]$, $\{t-2~,t,~t+2\}$, $\{t-3~,t,~t+3\}$  and the other two on $\{t\}$ contextual frames, respectively. \textcolor{black}{The TDNN layers are having a dilation of (1, 2, 3, 1, 1) respectively.} This is followed by a two-layered BiLSTM unit, mean and standard deviation pooling layer, three fully connected (FC) layers, and a softmax layer on top of the network that reveals the classification scores of stuttering disfluencies.  A ReLU nonlinearity activation function and a 1D batch normalization are applied after each layer except the statistical pooling layer. We apply dropout of 0.3. in FC layers. 
\par 
Consider an input speech sample with $T$ frames. The first five layers of \emph{StutterNet} focus on the small context of speech frames. For example, \textcolor{black}{layer 2 takes as input the sliced output} of layer 1 at time frames of $\{t-2,~t,~t+2\}$, which results in capturing a total temporal context of 9 with the help of the previous layer's context of $[t-2,~t+2]$. Similarly, layer 3 sees a total context of 15 time frames. The statistical pooling layer computes and concatenates the mean and standard deviation by aggregating all $T$ output frames at the end of BiLSTM output. The statistical pooling layer accumulates the information across the temporal dimension which makes it suitable for subsequent layers to operate on the entire temporal speech segment. The \emph{StutterNet} is trained to classify 5 different types of stuttering including \emph{core behaviors} and the fluent part of the PWS. For detailed \emph{StutterNet} architecture, please refer to~\cite{sheikh:hal-03227223}.

\begin{table}
\caption{\scshape StutterNet  Architecture, TC: Total Context, TDNN: Time Delay Neural Network Layer, FC: Fully Connected Layer, BILSTM: Bidirectinal Long Short-Term Memory (2 Layers), Layer Context of [t-2, t+2] means 5 Frames are Taken into Consideration two Before the Current Time Step and two After the Current Time Stamp.}
    \centering
    \footnotesize
    \renewcommand{\arraystretch}{1.2}
    \begin{tabular}{c|c|c|c}
         \hline
         Layer & Output Layer Size & Layer Context& TC\\
         \hline
         TDNN 1\ &  64 & $[t-2,~t+2]$&5\\
         TDNN 2 & 64& \{t-2,~t,~t+2\}&9\\
         TDNN 3 & 64 &\{t-3,~t,~t+3\}&15\\
         TDNN 4& 64 & \{t\}&15\\
         TDNN 5&  3*64 & \{t\}&15\\
        BiLSTM & 64 * 2 & \{t\} & $T$ \\
         Statistical Pooling& $3*64*2$$\times$1 &[0, $T$)&$T$\\
         FC1&$64$&-&$T$\\
         FC2&64&-&$T$\\
         FC3&NumClasses&-&{$T$}\\
         \hline
    \end{tabular}
    \vspace{-0.2cm}
    \label{tab:snet}
 \end{table}

\subsection{Limitations}
\vspace{-0.1cm}
Although this network has shown promising results on a SEP-28K dataset in SD, it has several deficiencies. First, it \textcolor{black}{does not} generalize quite well on unseen data and leads to overfitting due to the limited amount of data available for training. In addition to data scarcity, the SEP-28k dataset was collected from podcasts in a clean environment, which makes the trained \emph{StutterNet} \textcolor{black}{difficult} to generalize in other environmental conditions. Second, obtaining class-balanced datasets is extremely difficult and very expensive in the speech domain and the SEP-28k dataset is no exception. Deep neural network (DNN) classifiers trained on highly class-imbalanced datasets are usually biased towards the majority class, which results in poor modeling of minority classes~\cite{fernandez2018learning}. In addition to the above deficiencies, we found in our previous work \textcolor{black}{\emph{StutterNet}~\cite{sheikh:hal-03227223}} that the context is very important in SD. \textcolor{black}{Stuttering detection is a special type of speech characterization problem where each class (i.e., stuttering types) has a varying duration. For example, block lasts for a shorter duration than prolongations and repetitions. Therefore, a fixed length of context in the basic \emph{StutterNet} framework might not be the optimized one for detecting all types of stuttering.} A larger context increases the performance of prolongation and repetition types of disfluencies but decreases the recognition performance of fluent speech segments on the UCLASS dataset~\cite{sheikh:hal-03227223}.

\vspace{-0.1cm}
\section{Addressing Deficiencies}
\label{address}
\vspace{-0.1cm}
In this section, we address the three above-mentioned issues.

\vspace{-0.1cm}
\subsection{Class Imbalance}
\vspace{-0.1cm}
 
In class imbalance learning, the distribution of samples across each class is not uniform. The DNN classifiers, trained on highly imbalanced datasets generally perform poorly for the minority class and favor the majority class~\cite{krawczyk2016learning}. In the worst case, where there is an extreme imbalance in the training set, the majority class dominates the learning, and samples among the minority class may go undetected, thus affecting performance~\cite{imbalancedlearningfoundation}. 

\par 
\begin{figure}
    \centering
    \includegraphics[scale=0.33]{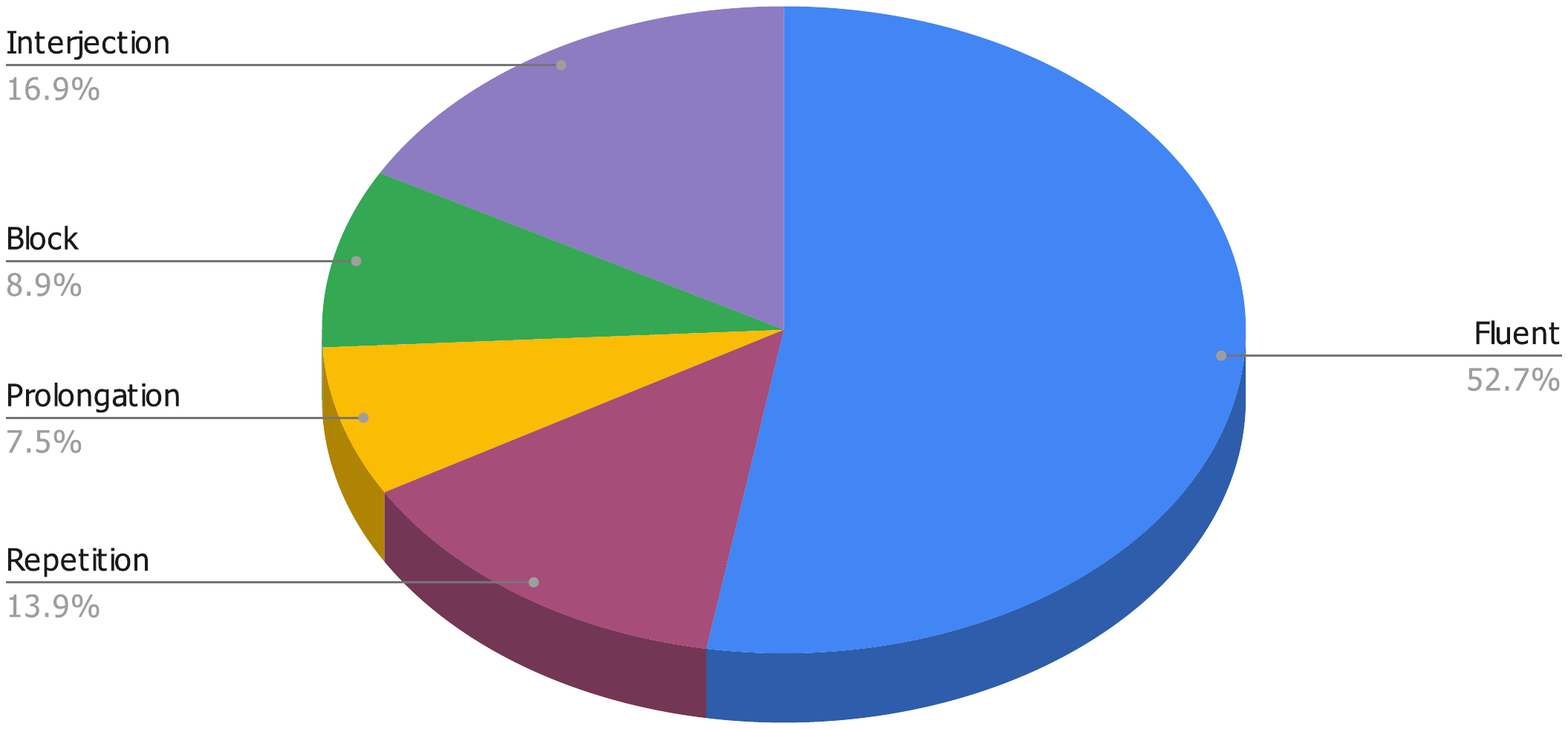}   
    \vspace{-0.2cm}
        \caption{Stuttering data distribution in SEP-28k dataset showing five classes consisting of single stuttering class.}
    \label{fig:dist}
    \vspace{-0.2cm}
\end{figure}

The class-imbalance is one of the major problems in real-world applications, including multi-lingual speech recognition~\cite{winata2020adapt}, and stuttering is no different as shown in Fig.~\ref{fig:dist}. In fact, stuttering is extremely imbalanced across fluent and other speech disfluencies. The Fig.~\ref{fig:dist}. shows that the interjections are the most common disfluency present in the SEP-28k dataset followed by repetitions, blocks, and prolongations and the overall distribution is approximately equivalent to the fluent distribution.

Collecting a balanced dataset is difficult and expensive for the stuttering detection task. Other datasets such as \textcolor{black}{Kassel State of Fluency (KSoF) (not publicly accessible)}~\cite{schuller2022acm}, FluencyBank~\cite{sep28k} also suffer from this issue.

\par 
Over the years, the class imbalance problem is one of the main concerns due to its prevalence, especially in the biomedical domain. Several methods have been proposed to tackle the class imbalance problem, which \textcolor{black}{are} mainly categorized into three groups: data-level, cost-level, and architecture-level~\cite{fernandez2018learning}. Data-level approaches attempt to re-balance the class distribution by means of some re-sampling methods which include: under-sampling, over-sampling, or combined over-sampling and under-sampling~\cite{fernandez2018learning, krawczyk2016learning}. The architecture-level approaches attempt to modify the existing algorithm or develop a new one to tune and adapt \textcolor{black}{it} for imbalanced datasets~\cite{krawczyk2016learning, fernandez2018learning}. Cost-level approaches attempt to influence the loss/objective function by providing comparatively higher misclassification cost penalties to the minority class in order to force the model to learn about minority classes~\cite{krawczyk2016learning, fernandez2018learning}. In DNNs, addressing the class-imbalance problem by re-sampling may either get rid of sensitive speech samples that are extremely important in training when under-sampling. It can also add numerous quantities of duplicated speech samples under the over-sampling strategy, which eventually makes the training expensive and makes the DNN model likely to overfit~\cite{fernandez2018learning}. Because of these limitations, we investigate cost-level and architecture-level approaches in this work.

\par 

For the cost-based approach, we modify the standard cross entropy loss by assigning weights to different classes\textcolor{black}{~\cite{classbalancedloss}}. We set the class weights inversely proportional to the number of samples. We define the weight for class $i$ as $w_i = \frac{N}{C * N_{i}}$ where $N$ is the number of training samples, $C$ is the number of classes, $N_i$ is the number of training samples for class $i$. Therefore, the weighted cross-entropy (WCE) over the train set can be defined as,

\begin{equation}
     \mathcal{L}_{\mathrm{WCE}} = \frac{1}{\mathcal{B}}\sum\limits_{b=1}^{\mathcal{B}}\frac{\sum\limits_{i}^{M} w_i * \log(p_i)}{\sum\limits_{i,~i \in \mathcal{B}}^{M} w_i}
     \label{eq:batchinv}
\end{equation}

where $\mathcal{B}$ is the number of batches, M is number of stuttered speech samples in a batch $b_i$, $p_i = \Big( \frac{e^{c_i}}{\sum_{j=1}^C e^{c_j}} \Big)$ is the predicted probability of class $c_i$ of sample $i$.

\par 


For architecture-level, we propose a multi-branched approach similar to the work by C. Lea \emph{et al.}~\cite{sep28k} and M. Bader-El-Den \emph{et al.}~\cite{classimbalance} to address the class imbalance issue in SD task.
Inspired by the fact that the number of fluent class samples is almost equal to the total number of samples in disfluent classes, we simultaneously classify fluent vs difluent classes in one output branch and subcategories of disfluent classes in another output branch. 
\textcolor{black}{The multi-branched architecture with two output branches is shown in Fig.~\ref{fig:mcsnet} (The figure is overall architecture of mulit-contextual \emph{StutterNet} with two contexts, For single contextual MB \emph{StutterNet}, only context = 5 is taken into consideration)}. This has one common encoder $\mathcal{E}$ ($\theta_{\mathrm{e}}$) section followed by two parallel branches referred as \emph{FluentBranch} $\mathcal{F}$ ($\theta_{\mathrm{f}}$) and \emph{DisfluentBranch} $\mathcal{D}$ ($\theta_{\mathrm{d}}$).
The embeddings from the encoder are processed with both the branches parallelly, where the  $\mathcal{F}$ is trained to distinguish between fluent and disfluent samples, and the  $\mathcal{D}$ is trained to differentiate within the disfluent sub-categories.

The objective is to optimize the sum of \emph{FluentBranch} loss $\mathcal{L}_\mathrm{f}$ and \emph{DisfluentBranch} loss $\mathcal{L}_\mathrm{d}$. For simplicity, a simple sum of the two losses has been taken into consideration and it works well. Thus, the overall objective function is define as:

\begin{equation}
    \mathcal{L}(\theta_{\mathrm{e}}, \theta_{\mathrm{f}}, \theta_{\mathrm{d}})  =  \mathcal{L}_\mathrm{f}(\theta_{\mathrm{e}}, \theta_{\mathrm{f}}) + \mathcal{L}_\mathrm{d}(\theta_{\mathrm{e}}, \theta_{\mathrm{d}}).
\end{equation}

During the evaluation step, if the \emph{FluentBranch} predicts the sample as fluent, then \emph{FluentBranch} predictions are considered otherwise \emph{DisfluentBranch} predictions are taken into consideration to reveal the stuttering category.

\vspace{-0.1cm}
\subsection{Data Augmentation} 
\vspace{-0.1cm}
Deep learning has achieved rapid progress in the domain of speech processing tasks including speech recognition~\cite{speechrecog}, speaker recognition~\cite{latentspeech}, emotion recognition~\cite{emotion}, speech disorder detection~\cite{sheikh:hal-03227223}. \textcolor{black}{However, as a drawback, deep learning based models are hungry for the data and require a substantial amount of annotated data for training, and \emph{StutterNet} is no exception. The model may require more stuttering data than other speech processing domains, as the presence of disfluencies in stuttering speech corpus is not frequent.}
\par 
Data augmentation is a popular technique that increases the quantity and diversity of the existing annotated training data, improves robustness, and avoids the overfitting of DNNs. For normal speech recognition, data augmentation demonstrates to be an effective approach for dealing with data scarcity and enhancing the performance of various DNN acoustic methods~\cite{Park2019}. Several data augmentation techniques have been investigated including pitch adjust~\cite{mixspeech}, spectral distortion~\cite{spdistort}, tempo perturbation~\cite{addnoise}, speed perturbation, cross-domain adaptation~\cite{crossdomain}, adding noise to clean speech~\cite{addnoise}, spectrogram deformation with frequency and time masking~\cite{Park2019}, mixspeech~\cite{mixspeech}, etc.
\par 
On the contrary, so far, very limited attention has been given to data augmentation targeting the speech disorder domain.  In~\cite{dadysarhtric}, speed and tempo perturbation based data augmentation were used to convert the normal speech to dysarthric speech impairment. In~\cite{jiao}, a voice conversion adversarial training based framework was used to simulate dysarthric speech from healthy speech. In~\cite{christensen2013combining}, normal speech samples (also called out-of-domain) were used as a data augmentation for dysarthric speech in the bottleneck feature extraction stage. In~\cite{xiong}, the speaker-dependent parameter was computed, which was then used for augmentation of scarce dysarthric speech in tempo adjustment. \textcolor{black}{In~\cite{woszczyk22_interspeech}, several data augmentation techniques such as noise, time stretching, pitch shifting, time shift, masking, etc., were analysed in Dementia detection}. There are some studies on data augmentation targeting \textcolor{black}{text based stuttering detection~\cite{Alharbi2018}, however, in the case of audio based stuttering/disfluency detection, this has not been studied and analysed deeply~\cite{w2v2da} }. 
\par
\begin{figure}
     \centering
      \begin{subfigure}[t]{0.3\textwidth}
         \centering
         \includegraphics[trim={0 0cm 0 2cm}, ,clip,width=160pt]{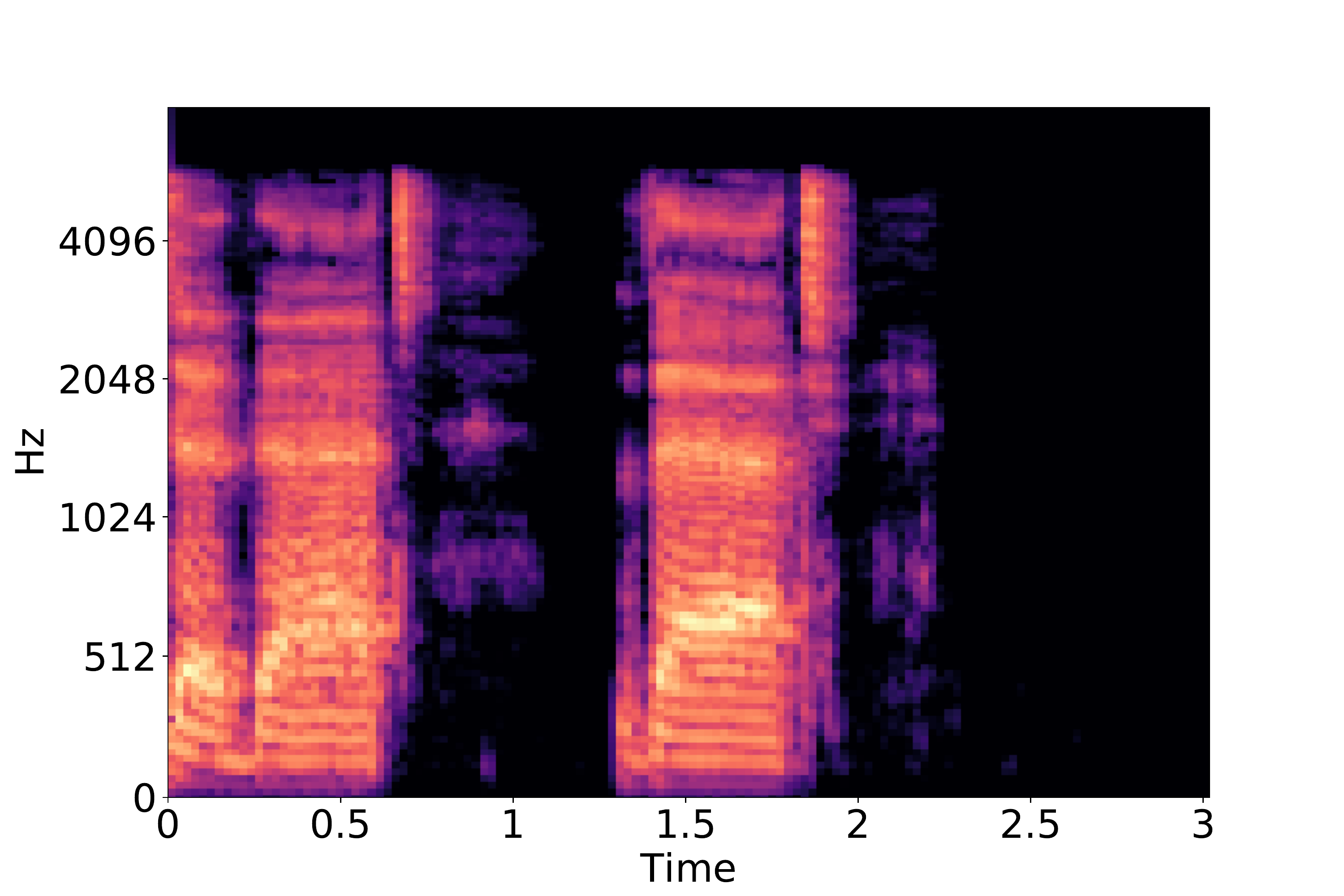}
         \caption{Repetition (clean speech)}   
         \label{fig:clean}
     \end{subfigure}
     \hfill
     \begin{subfigure}[t]{0.3\textwidth}
         \centering
         \includegraphics[trim={0 0cm 0 2cm}, ,clip,width=160pt]{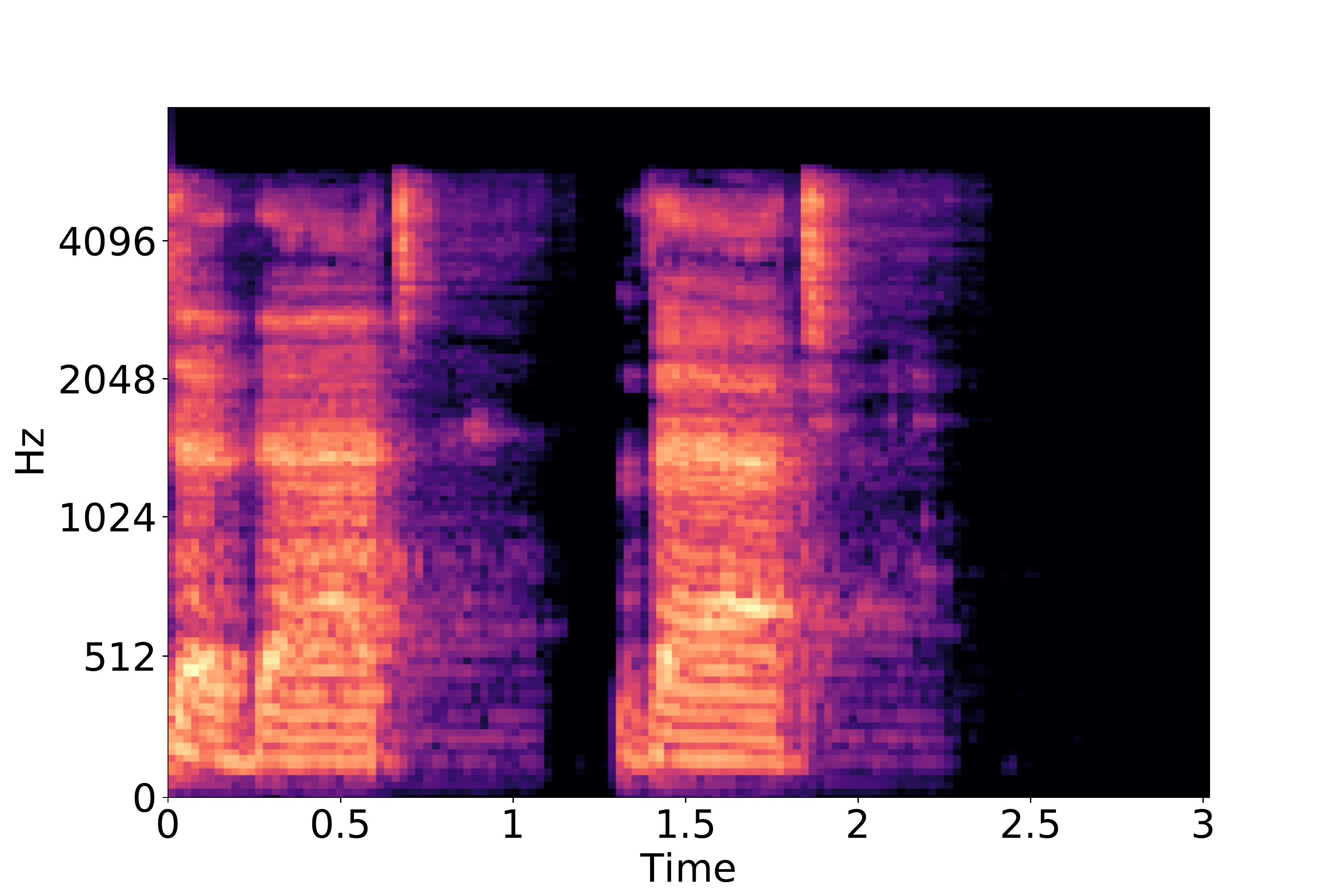}
         \caption{Repetition with reverberation}  
         \label{fig:reverb}
     \end{subfigure}
     \hfil
     \begin{subfigure}[t]{0.3\textwidth}
         \centering
         \includegraphics[trim={0 0cm 0 2cm}, ,clip,width=160pt]{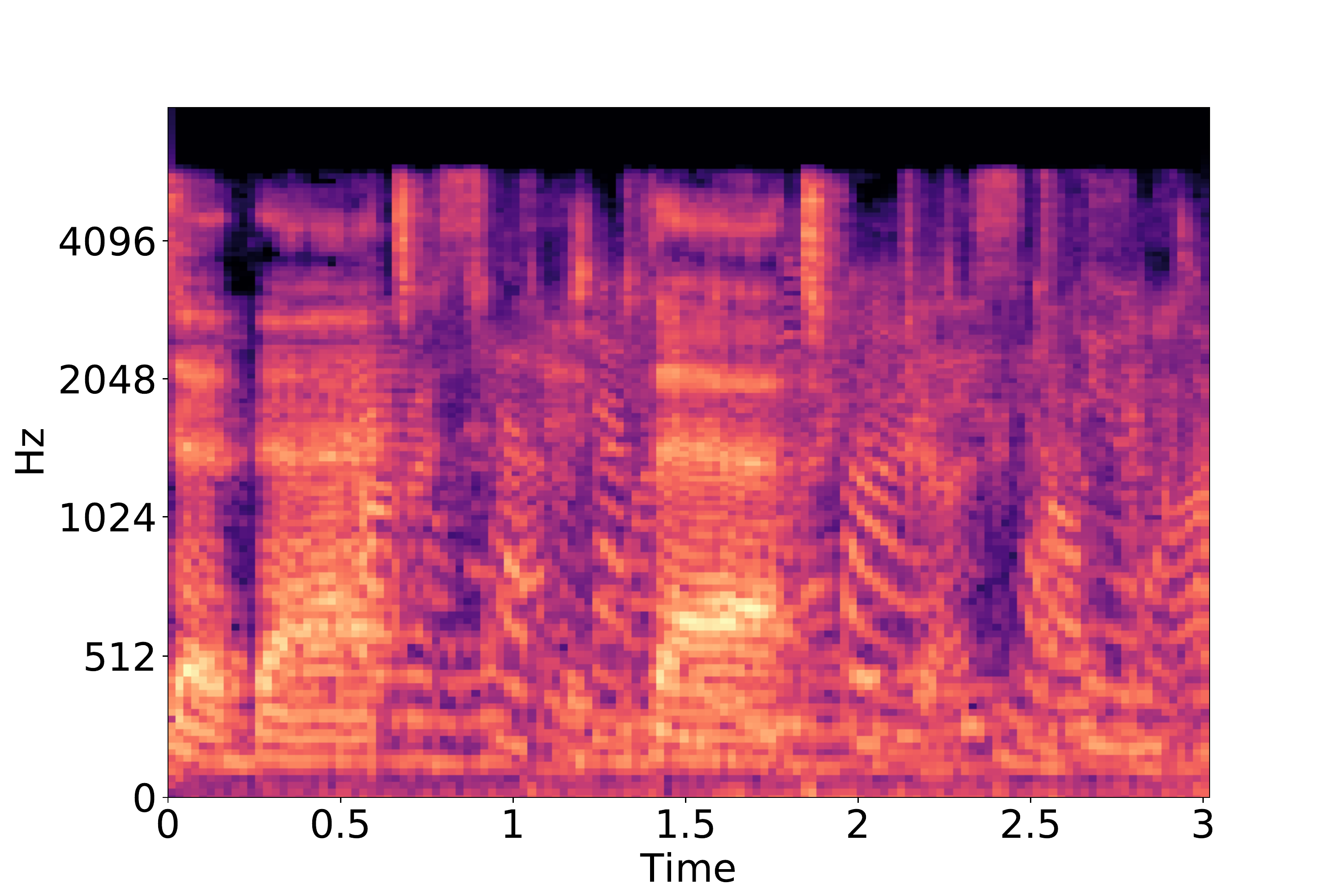}
         \caption{Repetition with babble noise}  
         \label{fig:bab}
     \end{subfigure}
     \hfill
     \begin{subfigure}[t]{0.3\textwidth}
         \centering
         \includegraphics[trim={0 0cm 0 2cm}, ,clip,width=160pt]{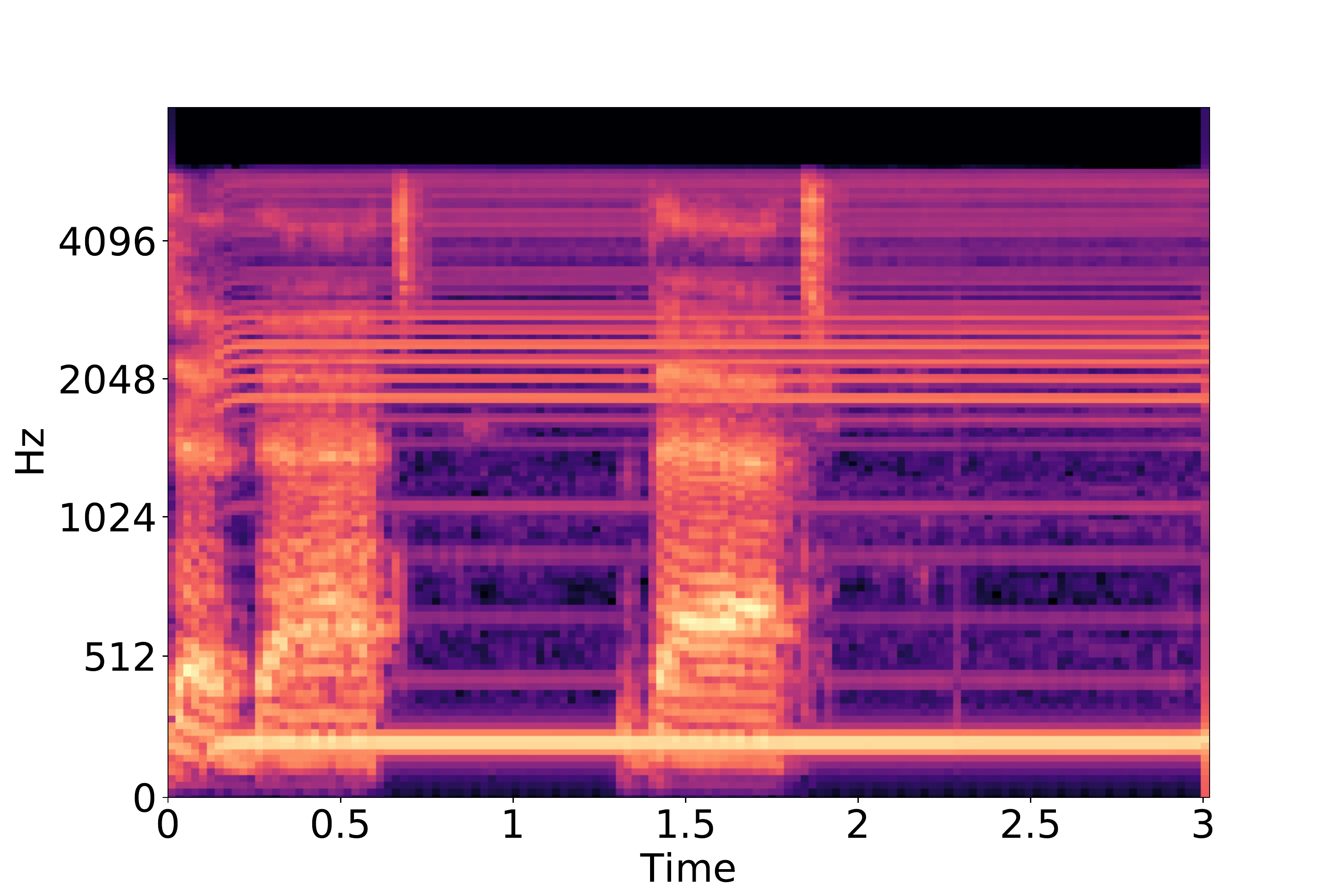}
         \caption{Repetition with white noise}  
         \label{fig:noise}
     \end{subfigure}
     \hfill
       \begin{subfigure}[t]{0.3\textwidth}
         \centering
         \includegraphics[trim={0 0cm 0 2cm}, ,clip,width=160pt]{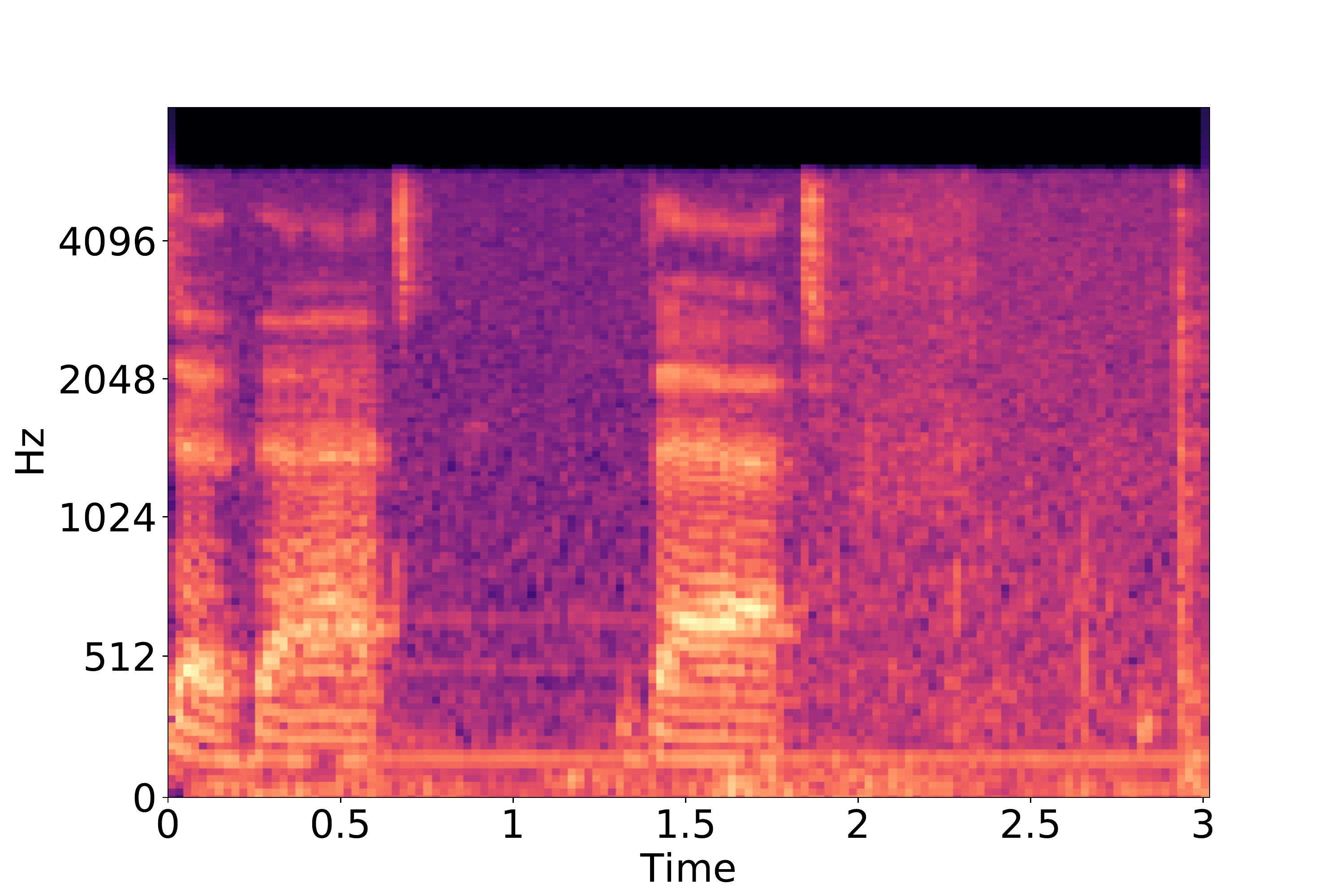}  
         \caption{Repetition with music}
         \label{fig:music}
     \end{subfigure}
        \caption{\textcolor{black}{Repetition stuttering with utterance \textit{``said that, that''} and the effect of various data augmentations (From 0 to 0.25 seconds, the speaker is saying \textit{``said''}, then followed by two repetitions \textit{``that''} from 0.25 to 0.6 and 1.4 to 1.75).}}
        \label{fig:dataugmentation}
\end{figure}

 
In the stuttering domain, the employment of data augmentation is not straightforward, because most of the data augmentations like time stretch, speed perturbation, etc, alter the underlying structure of the stuttering speech sample completely. Our approach employs reverberation and additive noises because it reflects the real-world scenario and does not change significantly the underlying stuttering in the speech sample as shown in Fig.~\ref{fig:dataugmentation}. Reverberation consists of convolving speech samples with room impulse responses. We utilize the simulated room impulse responses as described in~\cite{rir}. For additive noises, we utilize the MUSAN dataset, comprised of 60 hours of speech from 12 languages, 42 hours of music from various genres, and 900 hours of noises~\textcolor{black}{\cite{snyder2015musan}}.
\par 
\textcolor{black}{To augment the original speech samples, we combine the training "clean" set with below mentioned augmented copies which results in 5 times increase in training samples. }
\begin{enumerate}
   
    \item \emph{music:} A single music sample file randomly chosen from MUSAN is added to the original clean stuttering speech sample (SNR: 5-15dB) (The music sample file is trimmed or repeated as required to match the duration of the clean stuttered speech sample).
    \item \emph{noise:} Throughout the stuttered speech, samples from MUSAN noises are added at 1 sec intervals (SNR: 0-15dB).
     \item \emph{babble:} Speech samples from randomly three to seven speakers are summed together, then added to the original clean stuttered speech sample (SNR: 13-20dB).
    \item \emph{reverb:} The "clean" train set is convolved with simulated room impulse responses.  
\end{enumerate}
\textcolor{black}{All the data augmentation types shown in Fig.~\ref{fig:dataugmentation} were performed using Kaldi~\cite{Povey_ASRU2011} tool.}

\vspace{-0.1cm}
\subsection{Multi-contextual StutterNet} 
\vspace{-0.1cm}
The multi-contextual framework is based on the way humans perceive speech. In the cochlea, the input acoustic speech signal is partitioned into several frequency bands so that the information in each band can be filtered independently and thus processed in parallel in the human brain~\cite{allen1995humans}. 

Multi-contextual has been studied for action recognition in videos~\cite{twostream}, robust ASR~\cite{multistreamcnn, bourland, bourlard1996towards, hennansky, navid}, speech separation~\cite{zhangspeech}, where the input speech signal is processed in multiple streams/contexts (multiple time or frequency resolutions), etc. K.J Han \emph{et al}.~\cite{multistreamcnn} recently proposed a multi-stream\footnote{multi-stream, multi-scale, multi-resolution are the different names of multi-context.} convolutional neural network for robust acoustic modeling. Chiba~\emph{et al}.~\cite{Chiba2020MultiStreamAB} recently proposed multi-stream attention-based BiLSTM network for speech emotion recognition. Li~\emph{et al}.~\cite{li2018using} extracted deep features by training multi-stream hierarchical DNN for acoustic event detection. Moreover Sheikh \emph{et al.}~\cite{sheikh:hal-03227223} found that settings like context frame size optimized for one stuttering class are not good for other stuttering types. 
\par
\begin{figure*}
    \centering
    \includegraphics[scale=0.6]{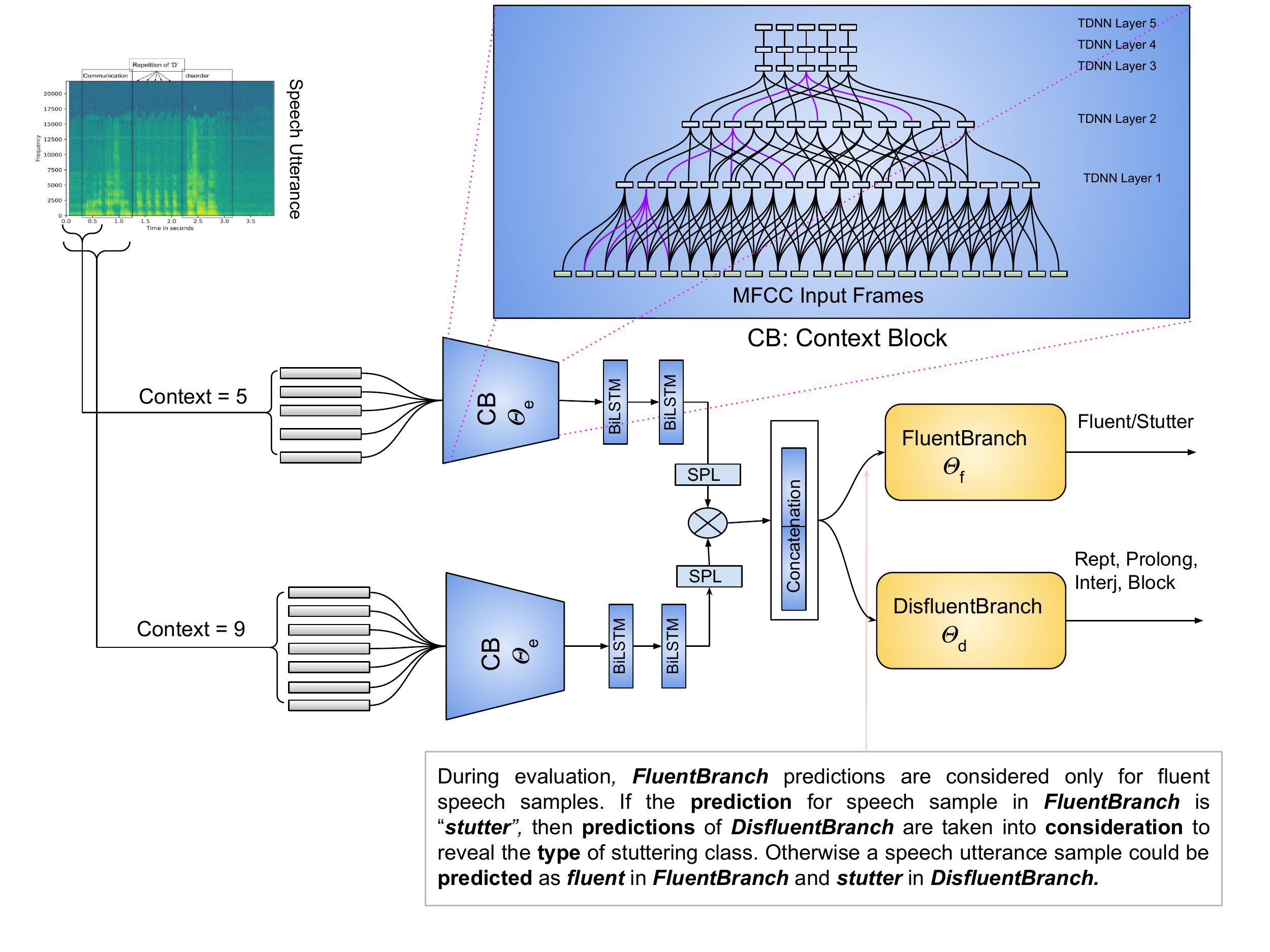}
    \vspace{-0.2cm}
    \caption{A schematic diagram of Multi-contextual \emph{StutterNet}, which is a multi-class classifier that exploits different variable contexts of (5, 9) in SD. The FluentBranch and DisfluentBranch are composed of 3 fully connected layers followed by a softmax layer for prediction of different stuttering classes, CB: Context Block, SPL: Statistical Pooling Layer. The context C (5, 9) here does not mean the TDNN layers, but rather it means the kernel-size which is number of frames taken into account when processing speech frames (C=5 means a context of 5 frames are taken at a time and C =9 means a context of 9 frames are taken at a time). These two contexts are jointly exploited in MC \emph{StutterNet} as shown in the left hand side of the figure. }
    \label{fig:mcsnet}
    \vspace{-0.2cm}
\end{figure*}

Exploiting this fact, we investigate how the multi-contextual neural networks will impact classification performance in the speech disorder domain, and in particular stuttering identification. In our preliminary study, we found that the context window improves the identification performance of two types of disfluencies on the \textcolor{black}{UCLASS dataset~\cite{sheikh:hal-03227223}}. As the context frame size increases in the \emph{StutterNet}, the performance detection of prolongation and repetition also increases, but decreases for fluent speech segments, and almost remains unaffected for a block type of stuttering. The prolongation and repetition last longer than other types of disfluencies. To address this issue, we exploit the variable contexts of MB \emph{StutterNet} by training the model jointly on different contexts as shown in Fig.~\ref{fig:mcsnet}. The pseudo-code of multi-contextual (MC) \emph{StutterNet} is provided in $Algorithm~1$. 

\begin{algorithm}
\footnotesize
\label{algo2}

\caption{Pseudo-code for MC \emph{StutterNet}}\label{alg:mcsnet}
\textbf{Output}: Predicted label set $\hat{y} \in (R, P, B, In, F)$ \\
\textbf{Input}: Stutter dataset with $\mathcal{D}  = (X_\mathrm{i}, f_\mathrm{i}, d_\mathrm{i}) $, where each sample $X_\mathrm{i}$ is $\mathcal{R}^{\mathrm{20 \times T}}$ MFCC input sequence \\
\begin{tabular}{l l}
$d$: Stutter label & $L$: Total loss\\
$f$: Pseudo label for FluentBranch & $b$: Sample batch\\ 
$K$: Number of epochs & $C$: Context \\
$CE$: Cross entropy loss function & $\lambda$: Learning rate\\ 
$p_\mathrm{f}$: Prediction of FluentBranch & $\theta = [\theta_\mathrm{e}, \theta_\mathrm{f}, \theta_\mathrm{d} ]$ : Network parameters\\
$p_\mathrm{d}$: Prediction of DisfluentBranch & $\theta_\mathrm{e}$ : Encoder parameters \\
$L_\mathrm{f}$: Loss of FluentBranch & $\theta_\mathrm{d}$ : DisfluentBranch parameters \\
$L_\mathrm{d}$: Loss of DisfluentBranch & $\theta_\mathrm{f}$ : FluentBranch parameters
\end{tabular}

\begin{algorithmic}[1]
\Ensure MC \emph{StutterNet} weight initialization 
    \For{ i in K}
    \For{ b in $\mathcal{D}$}
     \State \textbf{Forward Pass:}
     \State b = ApplyKaldiAugmentation(b)
     \State $out_\mathrm{5}$ = StutterNetBase(b, C = 5)  
     \State $out_\mathrm{9}$ = StutterNetBase(b, C = 9) 
     \State $stat_\mathrm{5}$ = StatisticalPooling($out_\mathrm{5}$)
     \State $stat_\mathrm{9}$ = StatisticalPooling($out_\mathrm{9}$)
     \State $stat_{\mathrm{embed}}$ = Concat($stat_5, stat_9$)
     \State $p_\mathrm{f}$ = FluentBranch($stat_{\mathrm{embed}}$)
     \State $p_\mathrm{d}$ =  DisfluentBranch($stat_{\mathrm{embed}}$)
     \State $L_\mathrm{f} = CE(p_\mathrm{f}, f_\mathrm{b})$ // $f_\mathrm{b}$:~Pseudo label of a batch
     \State $L_\mathrm{d} = CE(p_\mathrm{d}, d_\mathrm{b})$ // $d_\mathrm{b}$:~Stutter label of a batch
     \State $L(\theta_\mathrm{e}, \theta_\mathrm{f}, \theta_\mathrm{d}) = L_f(\theta_\mathrm{f}) + L_d(\theta_\mathrm{d})$
     \State \textbf{Backward Pass:}
     \State $\nabla \theta_\mathrm{b}$ = backward\_propagation($L(\theta)$, $\hat{y}$, $d$)
     \State \textbf{Parameter Updates:}
     \State $\theta_\mathrm{b} \gets \theta_\mathrm{b} - \lambda * \nabla \theta_\mathrm{b} $
    \EndFor
\State Compute CE loss on validation set
\If {$CE_{\mathrm{val}}$ $\leq$ $CE^{\mathrm{prev}}_{\mathrm{val}}$}
\If {patience $\leq$ 7}
\State Continue training $\rightarrow$ Go to step 2
\Else 
\State Stop training
\EndIf 
\EndIf

\EndFor

\end{algorithmic}
\end{algorithm}

\vspace{-0.1cm}
\section{Experimental setup}
\label{expts}
\vspace{-0.1cm}

We evaluate our proposed architecture thoroughly on the newly released SEP-28k stuttered dataset~\cite{sep28k}, and \textcolor{black}{for cross copora, we use the FluencyBank and LibriStutter datasets}. 

\subsection{Datasets}
\vspace{-0.1cm}
\emph{SEP-28k}: The SEP-28k stuttering dataset was curated from a set of  385 podcasts. \textcolor{black}{The original podcast recordings have varying lengths. From each podcast, 40 to 250 segments (each of length 3 seconds) are extracted which resulted in a total of 28,177 segments}. The original SEP-28k dataset contains two types of labels: stuttering and non-stuttering. The stuttering labels include blocks, prolongations, repetitions, interjections, and fluent segments, whereas the nonstuttering labels include unintelligible, unsure, no speech, poor audio quality, and music which are not relevant to our study. In our case study, we use only the stuttering single labeled samples. Out of 28,177 clips, we use only 23573 segments, among which 3286 are repetitions, 1770 are prolongations, 2103 are blocks and 12419 are fluent segments, and 3995 are interjections. This resulted in a total of 19.65 hours of data which includes 2.74 hours of repetition, 1.48 hours of prolongation, 1.75 hours of block, 10.35 hours of fluent speech, and 3.34 hours of interjections.
 After labeling, each 3-sec sliced speech segment is downsampled to 16~kHz. We randomly select 80\% of podcasts (without mixing podcasts) for training, 10\% of podcasts for validation, and the remaining 10\% of the podcast for evaluation in a 10-fold cross-validation scheme. The speaker information is missing from the SEP-28k dataset, so we divide the dataset based on \textcolor{black}{podcast identities} (assuming each podcast is having a unique speaker)\footnote{The details about the train, validation, and test set splits about the SEP-28k dataset is not publicly available. We create a protocol by ensuring no overlap of podcasts between train, validation, and test sets. The details are available in~\url{https://shakeel608.github.io/protocol.pdf}}.\\

\emph{FluencyBank}: The actual FluencyBank AudioVisual dataset was created by Nan Bernstein Ratner (University of Maryland) and Brian MacWhinney (Carnegie Mellon University) for the study
of fluency development. In our case study, we use the annotation done by the Apple similar to
SEP-28k [1]. This stuttering dataset was curated from 33 podcasts with 23 males and 10 females,
which resulted in a total of 4,144 segmented clips. Out of which, we only use 3355 samples (ignoring
the non-stuttering and multiple samples), among which 542 are repetitions, 222 are prolongations,
254 are blocks and 1584 are fluent segments, and 753 are Interjections. This results in a total of
2.80 hours of data which includes 0.45 hours of repetition, 0.19 hours of prolongation, 0.21 hours of block, 0.63 hours of interjection samples, and 1.32 hours of fluent samples. \textcolor{black}{We have considered those samples where at least two annotators agree with the same labeling of the segment.}\\

\textcolor{black}{\emph{Simulated LibriStutter}: The LibriStutter (English) consists of 50 speakers (23 males and 27 females), is approximately 20 hours and includes synthetic stutters for repetitions, prolongations, and interjections~\cite{libristutter}. Random stuttering was inserted within the four-second window of each speech signal. The original LibriStutter is having  6 classes including fluent, interjection\footnote{\url{https://borealisdata.ca/dataset.xhtml?persistentId=doi:10.5683/SP3/NKVOGQ}}, prolongation, sound, word and phrase repetitions. To make our experiments consistent with the SEP-28k dataset, we treat all repetitions as one class. After extracting samples based on the class label, we did not find any interjection samples in the dataset, so we only train with three classes including fluent, prolongation, and repetitions. For splitting of the dataset, please refer Table 3 in the response sheets.}

\vspace{-0.1cm}
\subsection{Training Setup}
\vspace{-0.1cm}
We implement models using the PyTorch library. The acoustic input features used in this case study are 20-dimensional MFCC features, which are generated every 10~ms using a 20~ms window and extracted using the \textcolor{black}{Librosa library~\cite{mcfee2015librosa}. For 3-dim pitch and phoneme features, we use Pykaldi~\cite{pykaldi} and Phonexia~\cite{phoenxia} tools respectively}. For training, we use Adam optimizer and cross-entropy loss function with a learning rate of $10^{-2}$ and batch size of 128. All the results reported in this paper are the average of the 10-fold validation technique and all the training experiments were stopped by early stopping criteria with patience of 7 on validation loss.

\subsection{Evaluation metrics}
To evaluate the model performance, we use the following metrics: macro F1-score and accuracy which are the standard and are widely used in the stuttered speech domain \cite{schuller2022acm, sheikh2021machine, kourkounakis2020detecting, sep28k, sheikh_acmmm}. The macro F1-score ($\mathcal{F}_1$) (which combines the advantages of both precision and recall in a single metric unlike unweighted average recall which only takes recall into account) from equation~(\ref{eq:macrof1}) is often used in class imbalance scenarios with the intention to give equal importance to frequent and infrequent classes, and also is more robust towards the error type distribution~\cite{opitz2019macro}.
\begin{equation}
    \mathcal{F}_1 = 1/C\sum_{k}F1_k = 1/C\sum_{k} \frac{2.P_kR_k}{P_k + R_k}
    \label{eq:macrof1}
\end{equation}
where C is the number of classes and $P_k$, $R_k$, and $F1_k$ denotes the precision, recall, and F1-score with respect to class $k$. 

\subsection{Experiments}
\textcolor{black}{This sections describes briefly the experiments carried out in this paper.}
\begin{itemize}
    \item \textcolor{black}{We carry out experiments using \emph{StutterNet}  with the two state-of-the art SD baselines including ResNet+BiLSTM and ConvLSTM models in the same settings to have a fair comparison.}
    \item \textcolor{black}{We perform experiments using weighted loss and multi-branched training schemes for addressing class-imbalance problem in stuttering domain, and also exploit the advantage of both the schemes in freezing parts of network.}
    \item \textcolor{black}{We experiment with the data augmented training to evaluate its performance in stuttering detection.}
    \item \textcolor{black}{We experiment with MC \emph{StutterNet} on top of data augmentation, and also analyse its performance in cross-corpora settings.}
\end{itemize}


\begin{table*}
\caption{{\scshape \textcolor{black}{Results with Baselines (BL) and Using Class Imbalance Learning (Clean Training)  (B: Block, F: Fluent, R: Repetition, P: Prolongation, In: Interjection, TA: Total Accuracy,  $\mathcal{F}_1$: Macro F1- Score), F$_{\mathrm{MFCC}}$: MFCC Input Features, F$_{\mathrm{F0}}$: 3-Dim (Pitch, Pitch-Delta, Voicing) Features, F$_{\mathrm{phone}}$: Phoneme Features,  MB: Multi Branch, WCE: Weighted Cross Entropy, $\mathcal{M}_{\mathrm{enc}}^{\mathrm{\mathrm{frz}}}$:Freezing Encoder, $\mathcal{M}_{\mathrm{enc, disf}}^{\mathrm{\mathrm{frz}}}$: Freezing Encoder and DisfluentBranch},  \textcolor{black}{$\mathcal{M}_{\mathrm{enc, fluent}}^{\mathrm{\mathrm{frz}}}$: Freezing Encoder and FluentBranch}.}}
\label{tab:Accuracy_cls_imb}
\centering
\renewcommand{\arraystretch}{1.2}
\begin{tabular}{c c}
 {      
 \scalebox{1}{\begin{tabular}{*{8}{c}}
    
     \hline
    \multicolumn{1}{c}{}&\multicolumn{5}{c}{Accuracy}&\multicolumn{1}{c}{}&\multicolumn{1}{c}{}\\
  \hline
    Method&R&P&B&In&F&TA&$\mathcal{F}_1$(\%)\\
       \hline
&&\multicolumn{6}{c}{\textcolor{black}{Baselines}}\\
 \hline
\textcolor{black}{
ConvLSTM  + F$_{\mathrm{MFCC}}$ (BL1) ~\cite{sep28k}}&22.83	&10.61&	06.34&	56.74&	72.35&	52.68&	34.00\\
\textcolor{black}{
ConvLSTM + F$_{\mathrm{phone}}$ (BL2) ~\cite{sep28k}}&10.18&	01.06&	00.35&	43.88&	74.48&	48.43& 24.00 \\
\textcolor{black}{
ConvLSTM + F$_{\mathrm{F0+MFCC}}$ (BL3) ~\cite{sep28k}}&19.28&	09.55&	08.51&	51.78&	66.60&	48.47& 30.80\\
\textcolor{black}{
 ResNet+BiLSTM (BL4)~\cite{kourkounakis2020detecting}} &	18.76&	41.24&	5.47&	57.18&	88.19&	62.36&	43.12\\
 
       StutterNet (BL5)~\cite{sheikh:hal-03227223}&27.14&	32.55&	2.96&	57.74&	87.60&	62.57&	42.84 \\
      \hline
      &&\multicolumn{6}{c}{\textcolor{black}{Class Imbalance}}\\
       \hline

        \textcolor{black}{
        ResNet+BiLSTM + WCE ~\cite{kourkounakis2020detecting}}&  28.90&	64.89&	33.79&	63.03&	46.90&	47.42&	41.00\\
\textcolor{black}{
        MB ResNet+BiLSTM ~\cite{kourkounakis2020detecting}}&  34.79&	30.19&	5.92&	49.26&	75.47&	55.62&	39.20\\ 
      
        StutterNet  + WCE (\emph{StutterNet}$_{\mathrm{WCE}}$) &  36.23	 & 59.73 & 	38.05 & 	61.19 & 	41.59 & 	45.26 & 	41.02 \\
   
            MB StutterNet (\emph{StutterNet}$_{\mathrm{MB}}$) & 35.26	&32.76&	7.21&	56.04&77.27&	58.56&	42.26 \\
        \hline
          $\mathcal{M}_{\mathrm{enc}}^{\mathrm{\mathrm{frz}}}$ & 39.82&	37.91&	10.45&	60.57&	73.49&	58.58&	44.42 \\
  $\mathcal{M}_{\mathrm{enc, disf}}^{\mathrm{\mathrm{frz}}}$ & 29.25&	45.85&	18.11&	56.88&		74.49&	58.18&	44.80 \\
  \textcolor{black}{$\mathcal{M}_{\mathrm{enc, fluent}}^{\mathrm{frz}}$} &31.15&	27.62&	05.01&	57.64&		73.64&	55.83   &38.60 \\

\hline
\end{tabular}}
 }

\end{tabular}
\end{table*}

\begin{table*}
\caption{{\scshape \textcolor{black}{ Results Using Data Augmentation (B: Block, F: Fluent, R: Repetition, P: Prolongation, In: Interjection, TA: Total Accuracy, $\mathcal{F}_1$: Macro F1- Score), A4: Babble + Reverberation + Music + Noise Augmentation, MB: Multi branch, WCE: Weighted Cross Entropy, $\mathcal{M}_{\mathrm{enc}}^{\mathrm{\mathrm{frz}}}$:Freezing Encoder, $\mathcal{M}_{\mathrm{enc, disf}}^{\mathrm{\mathrm{frz}}}$: Freezing Encoder and DisfluentBranch}, \textcolor{black}{$\mathcal{M}_{\mathrm{enc, fluent}}^{\mathrm{\mathrm{frz}}}$: Freezing Encoder and FluentBranch}.}}
\label{tab:Accuracyda}
\centering
\renewcommand{\arraystretch}{1.1}
\begin{tabular}{c c}

 {
     \scalebox{1}{\begin{tabular}{*{8}{c}}

     \hline
    \multicolumn{1}{c}{}&\multicolumn{5}{c}{Accuracy}&\multicolumn{1}{c}{}&\multicolumn{1}{c}{}\\
  \hline
    Method&R&P&B&In&F&TA&$\mathcal{F}_1$(\%)\\
        \hline

StutterNet + A4 & 28.67& 	32.53& 	3.69& 	64.24& 	 	88.70& 	64.69& 	45.30 \\
StutterNet + WCE (\emph{StutterNet}$_{\mathrm{WCE}}$) + A4  & 43.06& 	61.43& 	38.18& 	65.33& 		43.73& 	48.30& 	44.34 \\
\hline
  MB ResNet+BiLSTM + A4 &  33.27&	32.85&	0.99&	65.95&		70.92&	55.75&	39.44 \\
     
         MB StutterNet (\emph{StutterNet}$_{\mathrm{MB}}$) + A4 &34.05&	33.93&	4.32&	68.74&		78.88&	61.35&	44.03\\
        \hline
          $\mathcal{M}_{\mathrm{enc}}^{\mathrm{\mathrm{frz}}}$ + A4& 28.87&	30.94&	4.26&	64.12&		85.29&	62.85&	44.06 \\
  $\mathcal{M}_{\mathrm{enc, disf}}^{\mathrm{\mathrm{frz}}}$ + A4& 27.63&	36.33&	11.52&	62.31&		83.05&	62.14&	45.76 \\
   \textcolor{black}{$\mathcal{M}_{\mathrm{enc, fluent}}^{\mathrm{\mathrm{frz}}}$ + A4} &31.19&	25.14&	04.48&	61.30&	77.56&	58.32&		39.80 \\

\hline
\end{tabular}}
}

\end{tabular}
\end{table*}

\section{Results}
\label{results}
In this section, we discuss the results with class balanced training, data augmentation, and multi-contextual learning. We propose two modifications to the vanilla \emph{StutterNet} to address the class imbalance issue and one of them involves the loss function on the SEP-28k dataset. We further present the results of cross-corpora experiments on the FluencyBank and LibriStutter datasets. Table~\ref{tab:Accuracy_cls_imb} shows the results of state-of-the-art baselines (1st part are baselines) and impact of class balanced training. Table~\ref{tab:Accuracyda} depicts the results using data augmentation on top of class balanced training. Table~\ref{tab:results_mc} shows the results using MC \emph{StutterNet}.

\paragraph*{Baselines}
\textcolor{black}{
In addition to our single branch \emph{StutterNet} baseline (BL5), we first implement the state-of-the-art ConvLSTM and ResNet+BiLSTM as our baseline models for comparison purposes in the same settings. Our baseline model \emph{StutterNet} is performing well in almost all the disfluent classes except blocks as compared to the baseline model used in SEP-28k paper~\cite{sep28k} ConvLSTM (MFCC features) (referred to as BL1), ConvLSTM (phoneme features) (referred to as BL2), and ConvLSTM (pitch+MFCC features) (referred to as BL3). In addition, we also use one more model i.e, ResNet+BiLSTM classifier (referred to as BL4) from Kourkounakis \emph{et al.}~\cite{kourkounakis2020detecting}. Comparing our single branch baseline \emph{StutterNet} to BL4, the model performs well only in interjection and prolongation classes as shown in Table \ref{tab:Accuracy_cls_imb}. For subsequent comparison, we select BL1 (best among BL1, BL2 and BL3) and BL4.} 
\subsection{Class Imbalance}

\begin{figure}

    \hspace{-0.5cm} 
 \includegraphics[scale=0.35]{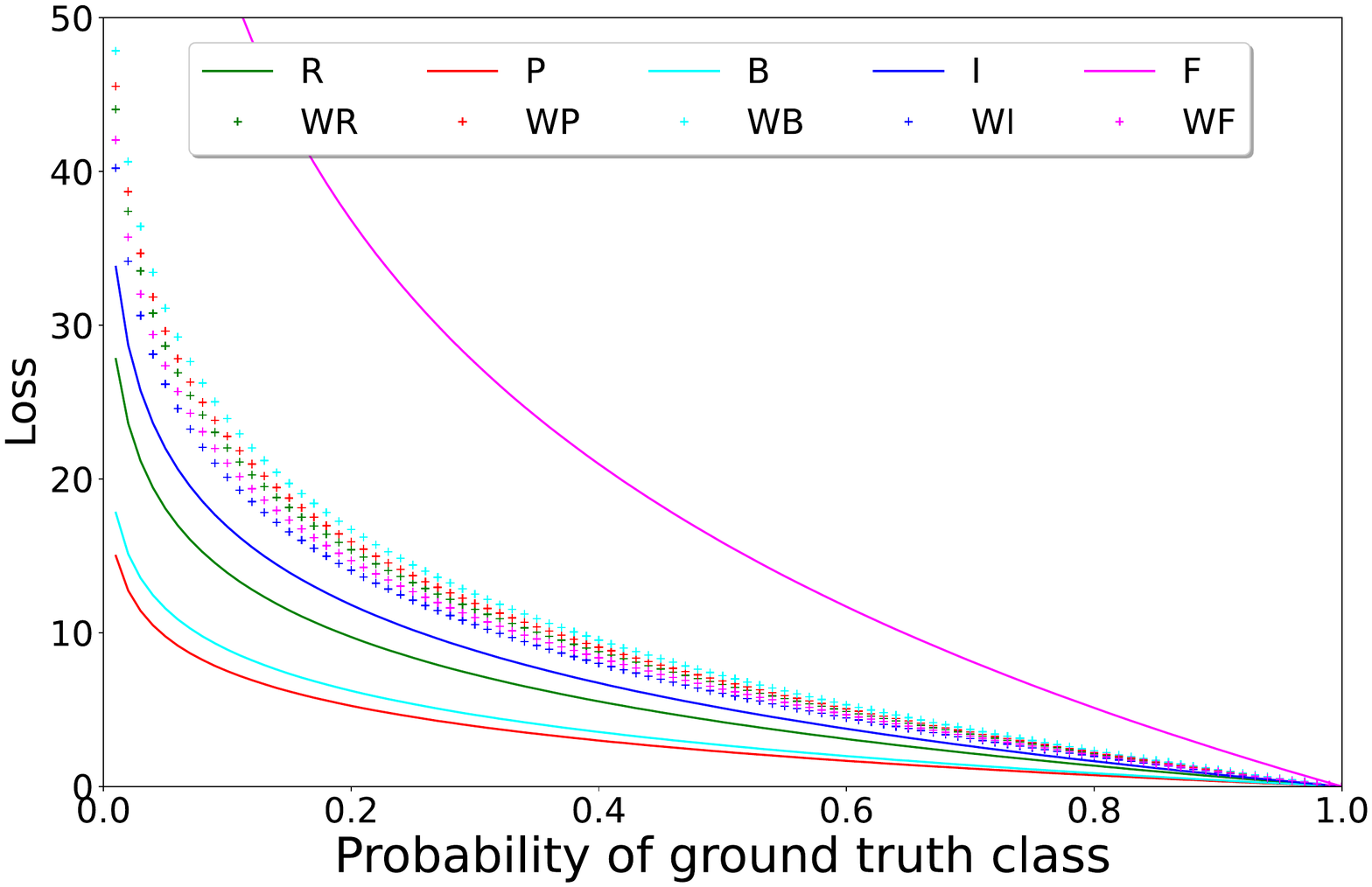}
    \caption{Class-wise loss values for different probabilities of ground truth classes. Here block, fluent, repetition, prolongation, and interjection classes are correspondingly denoted by B, F, R, P, and I. The standard cross entropy (CE) is shown by solid curves and its weighted version is shown by dashed ones (WX represents the weighted loss of class X).}
    \label{fig:wceloss}
\end{figure}

\subsubsection{Weighted cross entropy}

This section discusses the impact of applying weighted cross entropy to \emph{StutterNet} (which we call \emph{StutterNet}$_{\mathrm{WCE}}$) in order to improve the detection performance of minority classes including repetitions, blocks, and prolongations. Figure~\ref{fig:wceloss}. illustrates the class-wise training loss curves for normal and weighted cross-entropy loss functions. We can observe that for the standard cross entropy loss function, the majority class (i.e., fluents) exhibit higher loss values and it dominates the overall loss.

Therefore, during training with backpropagation, the number of updates for fluent class dominates the gradient values that in turn forces the model to focus mainly on correctly classifying/predicting the majority class. Thus, the minority classes including the blocks, prolongations, and repetitions are given less importance during training, which leads to their poor detection performance. Table \ref{tab:Accuracy_cls_imb} confirms that the detection performance of blocks, prolongations, and repetitions is very poor, as they are mostly predicted as fluent samples due to the class imbalance nature of the problem. The loss functions for the weighted cross entropy are shown by dashed curves in Fig.~\ref{fig:wceloss}. 
 \textcolor{black}{The figure indicates that applying weights to standard cross entropy loss function by equation (\ref{eq:batchinv}) forces the model to give balanced importance to each disfluency class while optimizing the parameters of the network during backpropagation.} This, in turn, helps in boosting the gradient updates for minority classes during training, and thus increases their detection performance in baselines including BL4, and BL5 as shown in Table \ref{tab:Accuracy_cls_imb}. The \emph{StutterNet$_\mathrm{WCE}$} gives a relative improvement of 33.49\%, 83.50\%, 1,185\%, and 5.98\% over \textcolor{black}{BL5 and 58.69\%, 462\%, 500\%, and 7.84\% over BL1}  for detecting repetitions, prolongations, blocks, and interjections, respectively. Table \ref{tab:Accuracy_cls_imb} also demonstrates the suitability of WCE with competitive ResNet+BiLSTM method \textcolor{black}{and results in 54\%, 57.35\%, 517\%, and 10.23\% relative improvement in repetitions, prolongations, blocks, and prolongations over BL4.}

\begin{figure*}[ht]
  \vspace{-1cm}
  \hspace{-2cm}
  \subfloat[\emph{StutterNet} (BL) ]{
	\begin{minipage}[c][1\width]{
	   0.3\textwidth}
	   \centering
	   \includegraphics[width=1.4\textwidth]{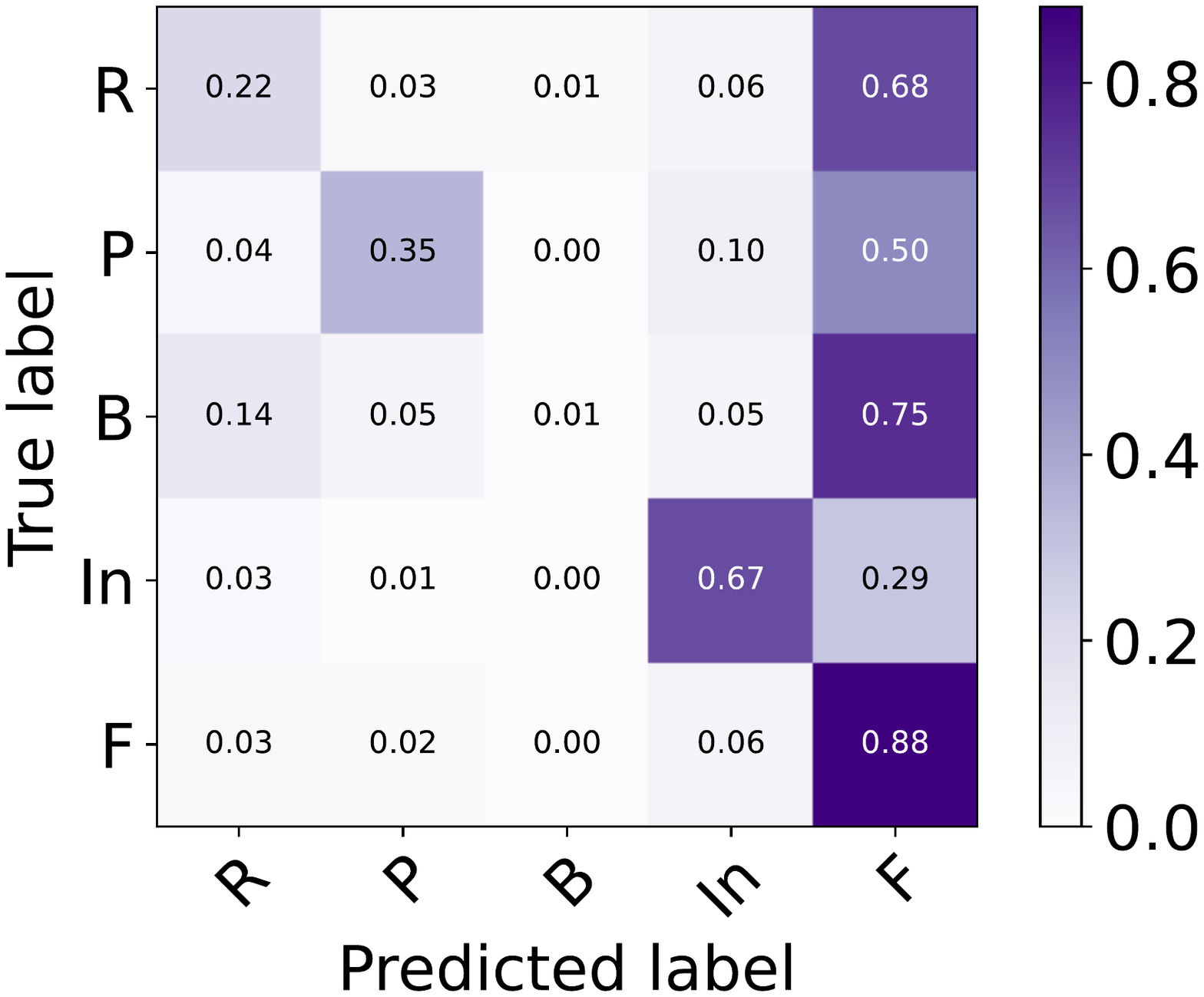}
          \vspace{-1.5cm}
	\end{minipage}}
 \hfill 	
  \subfloat[\emph{StutterNet} (WCE) (\emph{StutterNet}$_{\mathrm{WCE}}$)]{
	\begin{minipage}[c][1\width]{
	   0.3\textwidth}
	   \centering
	   \includegraphics[width=1.4\textwidth]{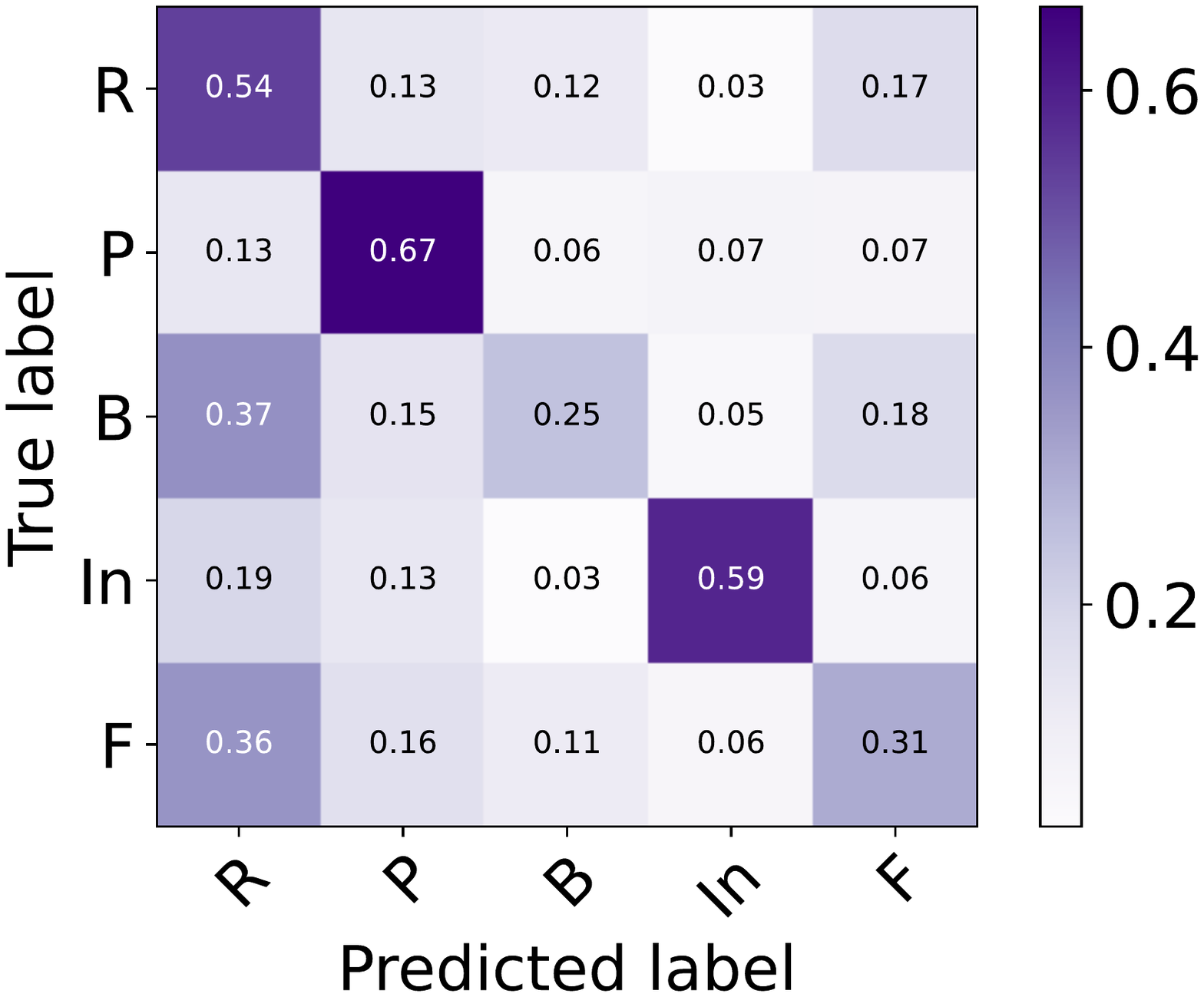}
          \vspace{-1.5cm}
	\end{minipage}}
 \hfill	
  \subfloat[MB \emph{StutterNet} (\emph{StutterNet}$_{\mathrm{MB}}$)]{
	\begin{minipage}[c][1\width]{
	   0.3\textwidth}
	   \centering
	   \includegraphics[width=1.4\textwidth]{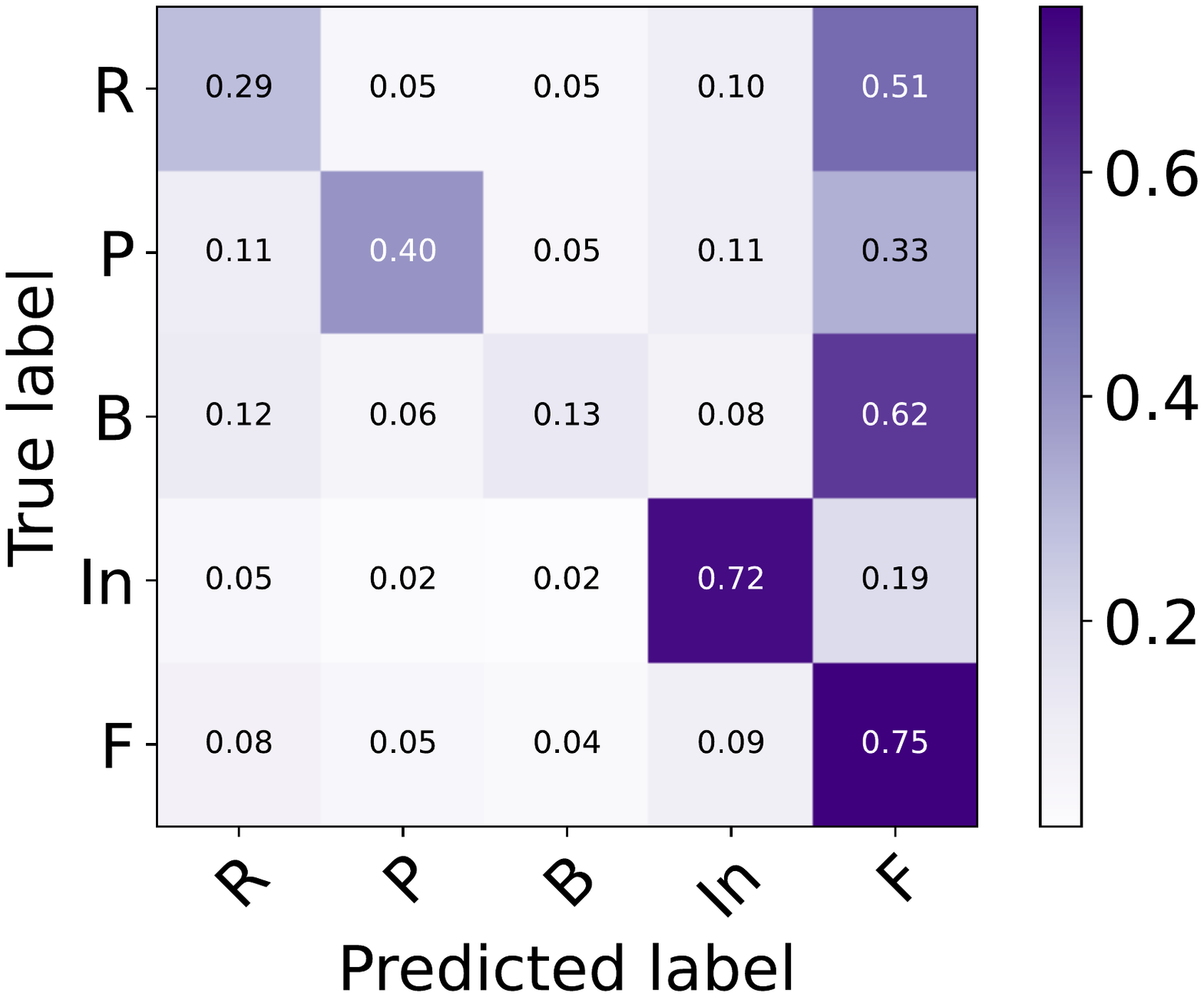}
          \vspace{-1.5cm}
	\end{minipage}}
\caption{Accuracy confusion matrices showing the confusion of fluent speech with repetitions and blocks. F: Fluent, R: Repetition, B: Block, P: Prolongation, In: Interjection, BL: Baseline, WCE: Weighted cross-entropy, MB: Multi-branch.}
 \label{fig:confusion}
\vspace{-0.2cm}
\end{figure*}

\par 
\subsubsection{Multi-branch training}
Moreover, we address the class imbalance problem via a multi-branch network (referred to as \emph{StutterNet}$_{\mathrm{MB}}$). This has two output branches: \emph{FluentBranch} and \emph{DisfluentBranch} as shown in Fig.~\ref{fig:mcsnet}. \textcolor{black}{(For \emph{StutterNet}$_{\mathrm{MB}}$, only a single context of five is taken into consideration) This method improves the detection performance of repetitions, prolongations, blocks by a relative margin of 29.92\%, 0.65\%, 144\% respectively over BL5, and 54.45\%, 208\%, 13.72\%, 6.80\% in repetitions, prolongations, blocks, and fluents respectively over BL1, and 88\%, 31.81\% in repetitions, blocks respectively over BL4, however, the BL4 is performing better in prolongations and interjection classes. We also applied multi-branch training in ResNet+BiLSTM which results in a relative improvement of 85.45\% in repetition class only as compared to the baseline BL4.  }
\par 
\textcolor{black}{In applying class balanced training, we found that there is drop in macro $\mathcal{F}_1$ score. Using \emph{StutterNet$\mathrm{WCE}$} and WCE based ResNet+BiLSTM, the macro $\mathcal{F}_1$ score drops from 42.84\% and 43.12\% to 41.02\% in \emph{StutterNet} and 41.02\% in ResNet+BiLSTM respectively. By employing multi-branch training, the macro $\mathcal{F}_1$ score drops very slightly from 42.84\% to 42.26\% in \emph{StutterNet} but it drops remarkably from 43.12\% to 39.20\% in ResNet+BiLSTM.}

\subsubsection{Analysis of confusion matrix}
 Using the WCE training scheme, the detection performance of minority classes improved remarkably at the cost of fluent accuracy. From the Fig.~\ref{fig:confusion} (a), we analyze that most of the repetitions and blocks are being classified as fluent speech. Initially, we hypothesize that it is most likely because of the imbalanced nature of the problem. Despite addressing the class imbalance problem in the stuttering domain using WCE and multi-branched training, we found that the block and repetition type of stuttering disfluencies are the ones that are still getting confused with the fluent class as depicted in Fig.~\ref{fig:confusion}~(b) and Fig.~\ref{fig:confusion}~(c). This makes intuitive sense because the blocks are closely related with fluent segments having just different initial silence or gasp followed by fluent utterances like fluent speech. The repetitions, on the other hand,  contain some word or phrasal repetitions, which are actually fluent utterances, if we carefully analyze their individual parts. Consider an utterance \emph{he he is a boy}. The word he is being repeated twice but is a fluent part if two \emph{he's} can be analyzed on an individual basis.

\subsubsection{Exploiting advantage of WCE and MB \emph{StutterNet}}
Since \emph{StutterNet}$_{\mathrm{WCE}}$ and \emph{StutterNet}$_{\mathrm{MB}}$ address the class-imbalance issue differently, we combine them to exploit both of their advantages. We first pre-train the \emph{StutterNet}$_\mathrm{WCE}$ and use it as a \emph{DisfluentBranch} in our multi-branched \emph{StutterNet}. After the pre-training step, we freeze parameters in two ways. First, we freeze the parameters of the contextual encoder only and fine-tune only the two output branches. We label this training scheme as $\mathcal{M}_{\mathrm{enc}}^{\mathrm{frz}}$. By exploiting this method, we achieve an overall detection improvement of 5.11\% in $\mathcal{F}_1$ over \emph{StutterNet}$_{\mathrm{MB}}$. Second, we freeze the base encoder and \emph{StutterNet}$_{\mathrm{WCE}}$ (\textit{DisfluentBranch}) and append with one more FluentBranch (to distinguish between fluents and stutter samples). We refer this as $\mathcal{M}_{\mathrm{enc, disf}}^{\mathrm{frz}}$ and it results in an overall improvement of 6.01\% in $\mathcal{F}_1$ over \emph{StutterNet}$_{\mathrm{MB}}$. 

\par 
 \textcolor{black}{We also experiment by first training the model using weighted cross entropy in \textit{FluentBranch}, and then by fine-tuning the \textit{DisfluentBranch} by freezing the parameters of the encoder and \textit{FlunetBranch}. We refer to it as $\mathcal{M}_{\mathrm{enc, fluent}}^{\mathrm{frz}}$ and the results for this configuration are shown in Table~\ref{tab:Accuracy_cls_imb} and \ref{tab:Accuracyda}. However, this training scheme degrades performance in almost all the disfluent classes in comparison to $\mathcal{M}_{\mathrm{enc}}^{\mathrm{frz}}$ (freezing encoder only) and $\mathcal{M}_{\mathrm{enc, disf}}^{\mathrm{frz}}$ (freezing encoder and \textit{DisfluentBranch}). This is possibly due to the reason that the \textit{base encoder} is trained only to distinguish between fluent and disfluent classes via \textit{FluentBranch}. Then freezing its parameters in a fine-tuning step further inhibits it more in learning sub-classes (repetitions, prolongations, interjections, and blocks) of the disfluent category which makes their overall detection performance lower.}

\subsection{Data augmentation}
The main experimental results obtained with various data augmentation techniques are shown in Table \ref{tab:Accuracyda}, where we compare the detection performance obtained with data augmented training to the baseline clean dataset. We first separately train the \emph{StutterNet}$_\mathrm{MB}$ on different data augmentation techniques which are described in Section~\ref{address}. We found that the training MB ResNet+BiLSTM and \emph{StutterNet}$_\mathrm{MB}$ with Kaldi augmentation increases the overall $\mathcal{F}_1$ performance. From Table~\ref{tab:Accuracyda}, it can be seen that the data augmentation does help in improving the $\mathcal{F}_1$ in almost all the cases. Applying data augmentation with a single branched and \emph{StutterNet}$_\mathrm{WCE}$, there is a relative improvement of 5.74\% and 3.50\% in $\mathcal{F}_1$ respectively, over the clean versions of the training. When data augmentation is applied to MB training, there is a relative improvement of 4.17\% and 0.61\% in $\mathcal{F}_1$ using \emph{StutterNet}$_\mathrm{MB}$ and ResNet+BiLSTM over clean training, respectively.

\par 
Moreover, applying Kaldi data augmentation on top of the $\mathcal{M}_{\mathrm{enc}}^{\mathrm{frz}}$ and $\mathcal{M}_{\mathrm{enc, disf}}^{\mathrm{frz}}$ training scheme, there is a relative improvement of 7.28\%, and 6.81\% in overall accuracy respectively. In addition to Kaldi augmentation, we also apply pitch scaling and bandpass filter as data augmentation, however, we did not achieve much improvement in SD. 

\par 

\begin{table*}
\caption{ {\scshape \textcolor{black}{Results Using MC \textit{StutterNet} with Clean and Data Augmentation. The MC \textit{StutterNet} also Contains MB Training. (B: Block, F: Fluent, R: Repetition, P: Prolongation, In: Interjection, TA: Total Accuracy, Bb: Babble, Rv:Reverberation, Mu:Music, No:Noise, A4: Bb + Mu + No + Rv Augmentation)}}}
 \label{tab:results_mc}
\centering
\begin{tabular}{c c}

 {
 \scalebox{1}{\begin{tabular}{*{8}{c}}
 
     \hline
    \multicolumn{1}{c}{}&\multicolumn{5}{c}{Accuracy}&\multicolumn{1}{c}{}&\multicolumn{1}{c}{}\\
   \hline
    Method&R&P&B&In&F&TA&$\mathcal{F}_1$(\%)\\
        \hline
    
           MC \emph{StutterNet}~(Clean)& 33.36&	37.15&	08.34&	59.63&		78.34	&59.77&	43.86\\

       \textcolor{black}{MC \emph{StutterNet}~+~Bb}&32.29	&37.11	&09.85	&66.50	&74.54	&58.71 & 44.40 \\
        \textcolor{black}{MC \emph{StutterNet}~+~Mu}  &32.08	&36.57	&07.98	&65.68	&77.67&	59.96  & 44.00 \\
        \textcolor{black}{MC \emph{StutterNet}~+~No} &28.54	&32.39	&04.75	&64.20	&80.42&	59.99 & 42.80  \\
        \textcolor{black}{MC \emph{StutterNet}~+~Rv} &31.98	&37.02	&05.83	&67.63	&77.41&	60.08  & 44.20 \\
        
        MC \emph{StutterNet}~+~A4& 34.65&	39.75&	10.12&	67.91&		77.89	&61.72& 46.00\\
\hline
\end{tabular}}
 }

\end{tabular}
\end{table*}

 \definecolor{lightpurple3}{rgb}{1.0, 0.55, 0.0} 
 \definecolor{lightpurple6}{rgb}{0.57, 0.63, 0.81}
 \definecolor{lightpurple8}{rgb}{0.53, 0.38, 0.56}
 \definecolor{lightpurple10}{rgb}{0.33, 0.41, 0.47}

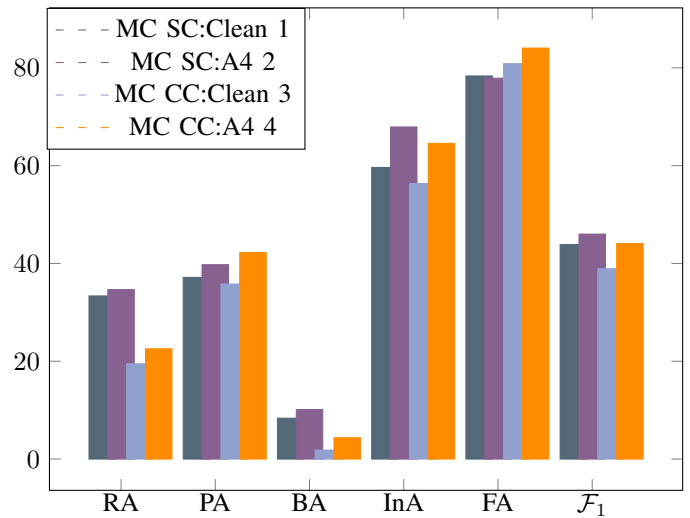
\begin{figure}
    \centering
    \begin{tikzpicture}
\begin{axis} [height=8cm,width=10cm,  
legend style={at={(0.2, 1)},anchor=north},
xtick={1, 6, 11, 16, 21, 26},
xticklabels={RA, PA, BA, InA, FA, $\mathcal{F}_1$},
cycle list name=exotic,
extra x tick style={tick label style={rotate=60}},
xticklabel style = {yshift=0.5ex}]
\addplot[ybar,fill=lightpurple10, draw=lightpurple10]coordinates {(0,33.36)};
\addlegendentry{MC SC:Clean 1}
\addplot[ybar,fill=lightpurple8, draw=lightpurple8]coordinates {(1,34.65)};
\addlegendentry{MC SC:A4 2}

\addplot[ybar,fill=lightpurple6, draw=lightpurple6]coordinates {(2,19.48)};
\addlegendentry{MC CC:Clean 3}
\addplot[ybar,fill=lightpurple3, draw=lightpurple3]coordinates {(3,22.54)};
\addlegendentry{MC CC:A4 4}

\addplot[ybar,fill=lightpurple10, draw=lightpurple10] coordinates {(5, 37.15) }; 
\addplot[ybar,fill=lightpurple8, draw=lightpurple8]coordinates {(6,39.75)};

\addplot[ybar,fill=lightpurple6, draw=lightpurple6] coordinates {(7, 35.80) }; 
\addplot[ybar,fill=lightpurple3, draw=lightpurple3]coordinates {(8, 42.22)};

 \addplot[ybar, fill=lightpurple10, draw=lightpurple10]coordinates {(10, 08.36) };
 \addplot[ybar,fill=lightpurple8, draw=lightpurple8]coordinates {(11, 10.12)};

 \addplot[ybar, fill=lightpurple6, draw=lightpurple6]coordinates {(12, 01.83) };
 \addplot[ybar,fill=lightpurple3, draw=lightpurple3]coordinates {(13, 04.36)};

\addplot[ybar, fill=lightpurple10, draw=lightpurple10]coordinates {(15, 59.63)};
\addplot[ybar,fill=lightpurple8, draw=lightpurple8]coordinates {(16, 67.91)};

\addplot[ybar, fill=lightpurple6, draw=lightpurple6]coordinates {(17, 56.36)};
\addplot[ybar,fill=lightpurple3, draw=lightpurple3]coordinates {(18,64.56)};

 \addplot[ybar, fill=lightpurple10, draw=lightpurple10]coordinates {(20, 78.34)};
 \addplot[ybar,fill=lightpurple8, draw=lightpurple8]coordinates {(21,77.89)};

\addplot[ybar,fill=lightpurple6, draw=lightpurple6] coordinates {(22, 80.85) }; 
\addplot[ybar,fill=lightpurple3, draw=lightpurple3]coordinates {(23, 84.04)};

 \addplot[ybar, fill=lightpurple10, draw=lightpurple10]coordinates {(25, 43.86)};
 \addplot[ybar,fill=lightpurple8, draw=lightpurple8]coordinates {(26, 46.00)};

\addplot[ybar,fill=lightpurple6, draw=lightpurple6] coordinates {(27, 38.92) }; 
\addplot[ybar,fill=lightpurple3, draw=lightpurple3]coordinates {(28, 44.07)};

\end{axis}
\end{tikzpicture}
    \caption{\textcolor{black}{Impact of data augmentation (A4) in cross corpora FluencyBank dataset with MC \emph{StutterNet} (R: Repetition, P: Prolongation, B: Block, In: Interjection, F: Fluent, XA: X: Disfluency Class and A: Accuracy, MC: Multi contextual \emph{StutterNet}, SC: Same corpora, CC: Cross corpora, A4: Augmentation). The bar plot clearly shows that the model MC \emph{StutterNet} trained on clean SEP-28k dataset fails to generalize on FluencyBank cross corpora data. Applying data augmentation improves the stuttering detection in cross domain corpora as shown orange bars (4$^{th}$ column in each disfluency).}}
    \label{fig:barplot}
    \vspace{-0.2cm}
\end{figure}


\subsection{Multi-contextual StutterNet}
Different contexts show different optimized class accuracies.
In order to exploit these different variable contexts and to improve the detection performance further, we propose a multi-contextual (MC) \emph{StutterNet} for SD as shown in Fig.~\ref{fig:mcsnet}. We jointly train and optimize the MC \emph{StutterNet} on the clean and augmented data using variable contexts of 5 and 9. The embeddings extracted from each context as depicted by $\mathcal{CB}$ block are passed to a two-layered BiLSTM unit, and then concatenated after applying statistical pooling layer (SPL), resulting in a $1\times2\times(2\times N)$-dimensional feature vector (where $N$ is layer size), which is then fed parallelly to two different branches including \emph{FluentBranch} and \emph{DisfluentBranch}  for class predictions. \textcolor{black}{This results in a relative improvement of 13.40\%, 15.67\%, 6.41\%, and 1.38\% in prolongation, block, interjection, and  fluent classes over \emph{StutterNet}$_{\mathrm{MB}}$ (clean), thus an improvement of  3.79\% in macro $\mathcal{F}_1$ score, however, employing multi-contextual training, we see a drop from 35.26\% to 33.36\% in repetition accuracy over the \emph{StutterNet}$_{\mathrm{MB}}$. In comparison to baseline BL5 (vanilla single branch \emph{StutterNet}), there is a relative improvement of 22.92\%, 14.13\%, 181.81\%, 3.27\% in repetition, prolongation, block, and interjection classes respectively. Thus a relative improvement of 2.38\% in macro $\mathcal{F}_1$ score. The MC \emph{StutterNet} also performs better in macro $\mathcal{F}_1$ score in comparison to state-of-the-art baselines BL1 and BL4. Applying data augmentation on top of the MC \emph{StutterNet}, we found that except noise augmentation, all other data augmentation types help in improving the macro $\mathcal{F}_1$ score in comparison to MC \emph{StutterNet} (clean)
The noise augmented samples help in improving the detection accuracy of fluent class. For interjections, we found that all the data augmentation helps, and for blocks, only  babble augmentation helps in improving their detection accuracy. We also found applying all four data augmentation techniques in MC \emph{StutterNet} results in an accuracy improvement in prolongation and repetition classes, however, with individual data augmentation, a drop in their accuracies can be observed in Table~\ref{tab:results_mc}. By applying all four data augmentation in MC \emph{StutterNet} training, there is an overall improvement of 1.76\%, 17.15\%, and 134\% in repetitions, prolongations, and blocks respectively as shown in the Table~\ref{tab:results_mc}, which is a 4.48\% relative gain in macro $\mathcal{F}_1$ score over the augmented \emph{StutterNet}$_{\mathrm{MB}}$ training.}

\subsection{Summary of proposed methods}
\vspace{-0.1cm}
This work advances the basic \emph{StutterNet} by addressing its limitations with three modifications. In Table~\ref{tab:results_summary}, we present a summary of the results demonstrating systematic improvements. We observe that all the proposed modifications help to gradually improve the performance and we achieve 7.37\% overall relative improvement in terms of macro $\mathcal{F}_1$ score.

\begin{table}
\caption{ {\scshape Summary of Results of Proposed Methods (A4: All Four Augmentation (Bb + Mu + No + Rv)}}
 \label{tab:results_summary}
\centering
\begin{tabular}{c c}
 {
 \scalebox{1}{\begin{tabular}{*{2}{c}}
    
   \hline
    Method&$\mathcal{F}_1$ (\%) \\
        \hline
StutterNet (BL5) (Clean)~\cite{sheikh:hal-03227223} & 42.84 \\
\hline
\multicolumn{2}{c}{\textcolor{black}{Class Imbalance}}\\
\hline
\emph{StutterNet}$_{\mathrm{WCE}}$ (Clean)& 41.02 \\
\emph{StutterNet}$_{\mathrm{MB}}$ (Clean)& 42.26 \\
 $\mathcal{M}_{\mathrm{enc}}^{\mathrm{frz}}$ (Clean)& 44.42 \\
$\mathcal{M}_{\mathrm{enc, disf}}^{\mathrm{frz}}$ (Clean)& 44.80 \\
\hline
 \multicolumn{2}{c}{\textcolor{black}{Data Augmentation}}\\
\hline
StutterNet + A4& 45.30\\
\emph{StutterNet}$_{\mathrm{WCE}}$ + A4& 44.34\\
 \emph{StutterNet}$_{\mathrm{MB}}$ + A4& 44.03\\

$\mathcal{M}_{\mathrm{enc}}^{\mathrm{frz}}$ + A4& 44.06\\
$\mathcal{M}_{\mathrm{enc, disf}}^{\mathrm{frz}}$ + A4& 45.76\\
\hline
 \multicolumn{2}{c}{\textcolor{black}{Multi-Contextual}}\\
\hline
MC StutterNet (Clean)& 43.86\\
MC StutterNet + A4& \textbf{46.00}\\

\hline
\end{tabular}}
 }

\end{tabular}
\end{table}

\subsection{Cross corpora evaluation}
\begin{table*}
\caption{ {\scshape \textcolor{black}{Results on Cross Corpora FluencyBank and Simulated LibriStutter Datasets. The First Row is From Table~\ref{tab:results_mc} Which Shows the Results on the Same Corpora SEP-28K Dataset. The Last Two Rows in the first half of the Table Show the Results, Where the Model is Trained on SEP-28k and Tested on FluencyBank in Cross Corpora Setting. (B: Block, F: Fluent, R: Repetition, P: Prolongation, In: Interjection), Bb: Babble, Rv: Reverberation, Mu: Music, No: Noise,  A4: All Four Bb + Rv + Mu + No Augmentation. The Multi-branched Training is also Considered in MC \emph{StutterNet}.}}}
\label{tab:ccres}
\centering
\begin{tabular}{c c}
  {
  \scalebox{1}{\begin{tabular}{*{10}{c}}
     \hline
    \multicolumn{1}{c}{}&\multicolumn{1}{c}{}&\multicolumn{1}{c}{}&\multicolumn{5}{c}{Accuracy}&\multicolumn{1}{c}{}&\multicolumn{1}{c}{}\\
    \hline
     Method&TrainSet&TestSet&R&P&B&In&F&TA&$\mathcal{F}_1$(\%)\\
 \hline   
             MC \emph{StutterNet}&SEP-28k&SEP-28k& 33.36&	37.15&	08.34&	59.63&		78.34	&59.77&	43.86\\

MC \emph{StutterNet} (Clean)&SEP-28k&FluencyBank & 19.48	&35.80&	01.83&	56.36&		80.85	&56.48	&38.92 \\
MC \emph{StutterNet}~+~A4&SEP-28k&FluencyBank& 22.54&	42.22&	4.36&	64.56&		84.04&	60.92&	44.07  \\
\hline

\textcolor{black}{MC \emph{StutterNet}} &\textcolor{black}{LibriStutter}&\textcolor{black}{LibriStutter}& 93.36&	76.26& NA&NA& 98.19&	96.11 & 91.00 \\
\textcolor{black}{MC \emph{StutterNet}(clean)} &\textcolor{black}{SEP-28k} &\textcolor{black}{LibriStutter}&00.65&	00.48&NA&NA&	99.95&	77.25&	30.00\\
\textcolor{black}{MC \emph{StutterNet}  + A4 }&\textcolor{black}{SEP-28k}&\textcolor{black}{LibriStutter}&01.11&	02.39&NA&NA&	97.35&	75.43&	31.00\\
\textcolor{black}{MC \emph{StutterNet}} &\textcolor{black}{LibriStutter}&\textcolor{black}{SEP-28k}&24.24&	55.11&NA&NA&	60.60&	53.21&	41.00\\ 
   
\hline
\end{tabular}}
 
 }

\end{tabular}
\end{table*}
Table~\ref{tab:ccres} shows the results on cross corpora datasets including FluencyBank and simulated LibriStutter. By optimizing on SEP-28k dataset in terms of data augmentation and multi-contextual, we aim to evaluate our proposed methodology MC \emph{StutterNet} on a cross corpora scenario. We train \emph{StutterNet} on SEP-28k dataset and evaluate it on \emph{FluencyBank} dataset which comprises samples from 33 podcasts. We found that the model trained on one corpus (SEP-28k) fails to generalize and performs poorly on cross-domain corpora. As can be seen from the Table \ref{tab:ccres} and Fig.~\ref{fig:barplot}, the $\mathcal{F}_1$ detection performance decreases remarkably from 43.86\% to 38.92\% employing clean training. \textcolor{black}{We hypothesize that the performance drop is due to the domain mismatch due to the difference in speaker accent and recording environment between the SEP-28k and FluencyBank datasets}. The repetitions and block classes show more degradation in their performance in cross corpora scenarios. Furthermore, we apply data augmentation in cross corpora evaluation, and we found that it boosts the detection performance of all the classes, which results in an improvement of 13.23\% in $\mathcal{F}_1$. 
\par 
\textcolor{black}{Experimental evaluation of MC \emph{StutterNet} on simulated LibriStutter dataset~\cite{libristutter} results in a macro $\mathcal{F}_1$ score of $\approx$ 91\% (shown in Table~\ref{tab:ccres}). However, the performance considerably drops when evaluated in cross-corpora setting in comparison to real stuttering dataset FluencyBank. The MC \emph{StutterNet} trained on SEP-28k dataset shows extremely poor performance when tested on simulated LibriStutter dataset. Applying data augmented training, we see there is minimal improvement of 1\% and 2\% in repetition and prolongation, respectively. The LibriStutter simulated dataset does not reflect the actual nature and characteristics of stuttered speech. The results also confirms that a model trained on simulated stuttered datasets should not be used in a real clinical condition.}

\definecolor{lightpurple1}{rgb}{0.6,0,0.6}
\definecolor{lightpurple2}{rgb}{0.6,0,0.6}
\definecolor{lightpurple3}{rgb}{0.0,0.7,1.3}
\definecolor{lightpurple4}{rgb}{0.0,0.7,1.3}
\definecolor{lightpurple5}{rgb}{0.0,0.7,1.3}
\definecolor{lightpurple6}{rgb}{0.0,0.7,1.3}
\definecolor{lightpurple7}{rgb}{0.0,0.7,1.3}
\definecolor{lightpurple8}{rgb}{0.0,0.7,1.3}
\definecolor{lightpurple9}{rgb}{0.0,0.7,1.3}
\definecolor{lightpurple10}{rgb}{0.0,0.7,1.3}
\definecolor{lightpurple11}{rgb}{0.0,0.7,1.3}
\definecolor{lightpurple12}{rgb}{0.0,0.7,1.3}
\definecolor{lightpurple13}{rgb}{0.8,0, 0}
\definecolor{lightpurple14}{rgb}{0.8,0, 0}

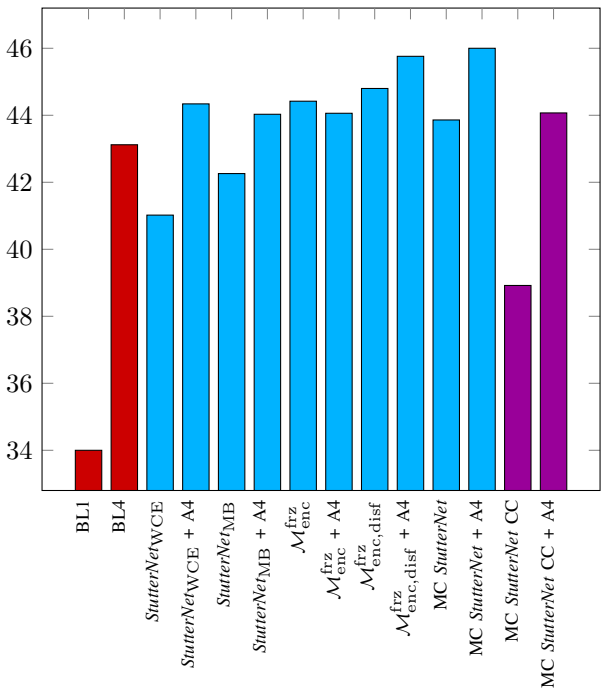
\begin{figure}
    \centering
    \begin{tikzpicture}
\begin{axis} [height=8cm,width=9cm,  
xtick=\empty,
cycle list name=exotic,
extra x ticks={0,1,2,3,4,5,6,7,8,9,10,11,12,13},
extra x tick labels ={BL1,	BL4,	\emph{StutterNet}$_{\mathrm{WCE}}$,	\emph{StutterNet}$_{\mathrm{WCE}}$ + A4,	\emph{StutterNet}$_\mathrm{MB}$,	\emph{StutterNet}$_{\mathrm{MB}}$ + A4,	$\mathcal{M}_{\mathrm{enc}}^{\mathrm{frz}}$,	$\mathcal{M}_{\mathrm{enc}}^{\mathrm{frz}}$ + A4,	$\mathcal{M}_{\mathrm{enc, disf}}^{\mathrm{frz}}$ ,	$\mathcal{M}_{\mathrm{enc, disf}}^{\mathrm{frz}}$ + A4,	MC \emph{StutterNet},	MC \emph{StutterNet} + A4,	MC \emph{StutterNet} CC,	MC \emph{StutterNet} CC + A4},
extra x tick style={tick label style={rotate=90}},
xticklabel style = {font=\scriptsize,yshift=0.5ex}]

\addplot[ybar,fill=lightpurple14]coordinates {(0,34)};

 \addplot[ybar,fill=lightpurple13] coordinates {   (1, 43.12) };
 
 \addplot[ybar, fill=lightpurple12]coordinates {   (2,41.02) };
 
\addplot[ybar, fill=lightpurple11]coordinates {    (3, 44.34)};

 \addplot[ybar, fill=lightpurple10]coordinates {   (4, 42.26)};
 
 \addplot[ybar, fill=lightpurple9]coordinates {   (5,44.03)};
 
 \addplot[ybar,fill=lightpurple8]coordinates {   (6,44.42)};
 
\addplot[ybar, fill=lightpurple7]coordinates {    (7,44.06)};

 \addplot[ybar, fill=lightpurple6]coordinates {   (8, 44.8)};
 
 \addplot[ybar, fill=lightpurple5]coordinates {   (9,45.76)};
 
 \addplot[ybar,  fill=lightpurple4]coordinates {   (10,43.86)};
 
 \addplot[ybar,  fill=lightpurple3]coordinates {   (11,46)};
 
 \addplot[ybar, fill=lightpurple2]coordinates {   (12,38.92)};
 
 \addplot[ybar, fill=lightpurple1]coordinates {   (13,44.07)};
\end{axis}
\end{tikzpicture}
    \caption{\textcolor{black}{Macro $\mathcal{F}_1$ score summary of proposed and baseline models (BL1 and BL4). The red, light blue and purple bars indicate the baseline, same-corpora setting, and cross-corpora settings. (CC: Cross Corpora with training on SEP-28k and evaluated on FluencyBank dataset, A4: All four data augmentation. The second last bar shows cross-corpora performance on FluencyBank when trained using a clean SEP-28k dataset and the last bar shows the same with data augmentation).}}
    \vspace{-0.2cm}
    \label{fig:graphmacrof1}
\end{figure}

\vspace{-0.1cm}
\section{Conclusion}
\label{conc}
\vspace{-0.1cm}

This paper addresses the problem of class imbalance in the stuttering domain. We address the class imbalance problem via two strategies including weighted cross entropy and a multi-branch training scheme. The weighted cross entropy loss function forces the \emph{StutterNet} classifier to give more attention to minority classes. We also investigate the effectiveness of data augmentation in the SD domain. For data augmentation, we employ reverberations and additive noises from the \textcolor{black}{MUSAN dataset~\cite{snyder2015musan}}. Additionally, we propose a MC \emph{StutterNet}, time delay based neural network for SD. The proposed \emph{MC StutterNet} is a multi-class classifier that exploits different variable contexts trained jointly on different contexts (5,~9) using CE. More importantly, the experiments using data augmentation over the FluencyBank dataset revealed that our methodology generalizes better in the cross corpora domain. For class imbalance, we have used only a simple weighting scheme in the cross-entropy loss function, which results in an accuracy trade-off between majority and minority classes. 
\par 
In general, the data augmentation helps in stuttering domain, however, the use of data augmentation in the stuttering domain is not straightforward, thus, is limited because most data augmentations, such as time stretch, speed perturbation, and so on, completely alter the underlying structure of the stuttering speech sample. In order for data augmentation to be more effective, a domain stuttering specific data augmentation is required to be developed. In addition, the stuttering detection domain has not matured enough, so a single metric like other speech domains which reflects the overall performance of a model is yet to be developed. In addition to accuracy metric, we have also used a macro F1-score ($\mathcal{F}_1$) which gives a good indication for the better evaluation of proposed methods. Moreover, we use joint training over multi contexts in this work, and it is possible that one context can dominate the training. \textcolor{black}{A visualisation summary of macro $\mathcal{F}_1$ score of models in stuttering detection is shown in Fig.~\ref{fig:graphmacrof1}}. 
\par 

The proposed methodology show promising results and it can detect if the stuttering is present in the speech sample or not, however, it cannot predict where exactly the stuttering occurs in the speech frames. For future study, we would like to explore combining the other different types of neural networks in SD to predict the frames where exactly the stuttering occurs. In addition to varying context, the investigation of varying depth and different number of convolutional kernels is also an interesting topic to study in SD. Moreover, the temporal information captured by recurrent neural networks can also be investigated in a multi-stream fashion for the identification of disfluent speech frames. 
\par 
\textcolor{black}{The performance comparison of the proposed systems with two state-of-the-art systems demonstrate that even though we achieve a noticeable advancements, the automated stuttering detection requires further research for developing a clinically usable system. The stuttering detection is fundamentally a challenging task due to inter-person variations, language/accent/dialect variability and other speaking variations. The scopes of the current work are limited to addressing basic problems related to stuttering detection. This work can be extended with speaker-adaptive training, domain adaptation to further improve the stuttering detection problem.}
\par 
\textcolor{black}{Regarding cognitive aspects of stuttering are well modeled or not as the explainability analysis of the used deep models along with auxiliary data (e.g., functional magnetic resonance imaging or fMRI data) is yet to be explored. However, we think the base architecture (i.e., MC \emph{StutterNet}) developed in this work could still be useful for such explainability analysis. The stuttering detection can be improved with multimodality by integrating with visual cues related to head nodding, lip tremors, unusual lip shapes, quick eye blinks, facial expressions, etc. We think that the same base acoustic model can be used in a fusion framework.}
\par 

\textcolor{black}{We have found that the blocks are the most difficult to detect. It would be interesting to analyze the disfluencies of the speakers which are hardest to identify by doing more ablation analysis. The future work includes exploring self-supervised models that exploit unlabelled audio data.}

\vspace{-0.1cm}
\section*{Acknowledgment}
\vspace{-0.1cm}
\footnotesize{
 This work was made with the support of the French National Research Agency, in the framework of the project ANR BENEPHIDIRE (18-CE36-0008-03). Experiments presented in this paper were carried out using the Grid’5000 testbed, supported by a scientific interest group hosted by Inria and including CNRS, RENATER and several universities as well as other organizations(see https://www.grid5000.fr) and using the EXPLOR centre, hosted by the University of Lorraine.}

\bibliographystyle{IEEEtran}
\balance
\bibliography{ref.bib}

\end{document}